\documentclass[11pt,a4paper,DIV=11,numbers=noenddot,parskip=half]{scrartcl}
\pdfoutput=1

\usepackage[T1]{fontenc}
\usepackage{lmodern}
\usepackage[utf8]{inputenc}
\usepackage{amsmath}
\usepackage{amssymb}
\usepackage{graphicx}
\usepackage{enumitem}
\usepackage[absolute]{textpos}
\usepackage[noadjust]{cite}
\usepackage[affil-it]{authblk}
\usepackage{xcolor}
\usepackage{tabularx}
\usepackage{booktabs}
\usepackage{multirow}
\usepackage{listings}
\usepackage{bold-extra}
\usepackage[font=small,labelfont=bf,format=plain,margin=0.05\textwidth]{caption}
\usepackage[pdftitle={Integral Reduction with Kira 2.0 and Finite Field Methods},
  pdfauthor={Jonas Klappert, Fabian Lange, Philipp Maierhoefer, and Johann Usovitsch},
  pdfkeywords={XXX},
  bookmarks=true, linktocpage,
  colorlinks=true, allbordercolors=white, allcolors=blue]{hyperref}

 \newcounter{notecount}
 
\newcommand{\code}[1]{\texttt{#1}}

\newcommand*{\kira}{\code{Kira}}
\newcommand*{\kiraZO}{\code{Kira\;2.0}}
\newcommand*{\pyred}{\code{pyRed}}
\newcommand*{\firefly}{\code{FireFly}}
\newcommand*{\flint}{\code{FLINT}}
\newcommand*{\mpi}{\code{MPI}}
\newcommand*{\git}{\code{Git}}
\newcommand*{\gitlab}{\code{GitLab}}
\newcommand*{\fermat}{\code{Fermat}}

\newcommand{\pheaderline}{{\footnotesize TTK-20-24\\P3H-20-041\\FR-PHENO-2020-11\\MITP/20-044}}

\title{Integral Reduction with Kira 2.0 and\\ Finite Field Methods}

\author[,a]{Jonas Klappert%
  \thanks{E-mail: \href{mailto:klappert@physik.rwth-aachen.de}{klappert@physik.rwth-aachen.de}}}

\author[,a]{Fabian Lange%
  \thanks{E-mail: \href{mailto:fabian.lange@kit.edu}{fabian.lange@kit.edu}}}

\author[,b]{Philipp Maierh\"ofer%
  \thanks{E-mail: \href{mailto:philipp.maierhoefer@physik.uni-freiburg.de}{philipp.maierhoefer@physik.uni-freiburg.de}}}

\author[,c]{Johann Usovitsch%
  \thanks{E-mail: \href{mailto:jusovits@uni-mainz.de}{jusovits@uni-mainz.de}}}

\affil[a]{Institute~for~Theoretical~Particle~Physics~and~Cosmology,
          RWTH~Aachen~University, 52056~Aachen, Germany}

\affil[b]{Physikalisches~Institut, Albert-Ludwigs-Universit\"at~Freiburg,
          79104~Freiburg, Germany}

\affil[c]{PRISMA~Cluster~of~Excellence, Institut~f\"ur~Physik,
          Johannes~Gutenberg-Universit\"at~Mainz, 55099~Mainz, Germany}

\date{}

\begin{document}

\maketitle
\thispagestyle{empty}

\begin{abstract}
  We present the new version \code{2.0} of the Feynman integral reduction program \kira{} and describe the new features.
  The primary new feature is the reconstruction of the final coefficients in integration-by-parts reductions by means of finite field methods with the help of \firefly{}.
  This procedure can be parallelized on computer clusters with \mpi{}.
  Furthermore, the support for user-provided systems of equations has been significantly improved.
  This mode provides the flexibility to integrate \kira{} into projects that employ specialized reduction formulas, direct reduction of amplitudes, or to problems involving linear system of equations not limited to relations among standard Feynman integrals.
  We show examples from state-of-the-art Feynman integral reduction problems and provide benchmarks of the new features, demonstrating significantly reduced main memory usage and improved performance w.r.t.\ previous versions of \kira{}.
\end{abstract}

Keywords: Feynman diagrams, multi-loop Feynman integrals, dimensional regularization, Laporta algorithm, modular arithmetic, computer algebra

\begin{textblock*}{11em}(\textwidth,17mm)
\raggedright\noindent
\pheaderline
\end{textblock*}

\newpage

\section*{NEW VERSION PROGRAM SUMMARY}

\textit{Program title:} \kira{}

\textit{Developer's repository link:} \url{https://gitlab.com/kira-pyred/kira}

\textit{Licensing provisions:} GNU General Public License 3 (GPL)

\textit{Programming language:} \texttt{C++}

\textit{Journal Reference of previous version:} P.~Maierhöfer, J.~Usovitsch and P.~Uwer, \emph{{Kira—A Feynman integral
  reduction program}},
  \href{https://doi.org/10.1016/j.cpc.2018.04.012}{\emph{Comput.\ Phys.\ Commun.}~{\bfseries 230}
  (2018) 99}
  [\href{https://arxiv.org/abs/1705.05610}{{\ttfamily 1705.05610}}].

\textit{Does the new version supersede the previous version?:} Yes.

\textit{Reasons for the new version:} Implementation of new features, some of which improve the performance significantly for many problems.

\textit{Summary of revisions:}
The primary new feature is the reconstruction of the final coefficients in integration-by-parts reductions by means of finite field methods with the help of \firefly{}~[1,2].
This procedure can be parallelized on computer clusters with \mpi{}.
Further improvements include the expanded support for user-provided systems of equations as well as a new feature to reduce the main memory usage when generating the system of equations.

\textit{Nature of problem:}
The reduction of Feynman integrals to a smaller set of master integrals is a central strategy for high precision calculations in theoretical particle physics, e.g.\ for cross sections.
Furthermore, the reduction is a key ingredient in many methods to calculate the master integrals themselves.

\textit{Solution method:}
\kira{} generates a system of equations employing integration-by-parts~[3,4] and Lorentz-invariance identities~[5] as well as symmetry relations.
Linearly dependent equations are removed and master integrals are identified by solving the system over a finite field.
The resulting system can be solved with two methods:
Since version 1.0, \kira{} can solve the system algebraically with the help of \fermat{}~[6].
New in this version is the strategy of solving the system many times over finite fields and reconstructing the master integral coefficients with the help of \firefly{}~[1,2].
Both strategies can also be applied to arbitrary homogeneous linear systems of equations.

\textit{References:}

[1] J.~Klappert and F.~Lange, \emph{{Reconstructing rational functions with
  FireFly}}, \href{https://doi.org/10.1016/j.cpc.2019.106951}{\emph{Comput.\
  Phys.\ Commun.}~{\bfseries 247} (2020) 106951}
  [\href{https://arxiv.org/abs/1904.00009}{{\ttfamily 1904.00009}}].

[2] J.~Klappert, S.~Y. Klein and F.~Lange, \emph{{Interpolation of dense and
  sparse rational functions and other improvements in FireFly}},
  \href{https://doi.org/10.1016/j.cpc.2021.107968}{\emph{Comput.\
  Phys.\ Commun.}~{\bfseries 264} (2021) 107968}
  \href{https://arxiv.org/abs/2004.01463}{{\ttfamily 2004.01463}}.

[3] F.~V. Tkachov, \emph{{A theorem on analytical calculability of 4-loop
  renormalization group functions}},
  \href{https://doi.org/10.1016/0370-2693(81)90288-4}{\emph{Phys.\ Lett.\ B}
  {\bfseries 100} (1981) 65}.

[4] K.~G. Chetyrkin and F.~V. Tkachov, \emph{{Integration by parts: The algorithm
  to calculate $\beta$-functions in 4 loops}},
  \href{https://doi.org/10.1016/0550-3213(81)90199-1}{\emph{Nucl.\ Phys.}~{\bfseries B192}
  (1981) 159}.

[5] T.~Gehrmann and E.~Remiddi, \emph{{Differential equations for two-loop
  four-point functions}},
  \href{https://doi.org/10.1016/S0550-3213(00)00223-6}{\emph{Nucl.\ Phys.}~{\bfseries B580}
  (2000) 485}
  [\href{https://arxiv.org/abs/hep-ph/9912329}{{\ttfamily hep-ph/9912329}}].

[6] R.~H. Lewis, \emph{{Computer Algebra System Fermat}},
  \href{https://home.bway.net/lewis}{https://home.bway.net/lewis}.

\newpage

\pagenumbering{Alph}
\clearpage

\tableofcontents
\pagenumbering{roman}

\clearpage

\pagenumbering{arabic}

\section{Introduction}

At the Large Hadron Collider, the high energy regime of the Standard Model is probed with ever increasing accuracy.
To meet the precision requirements for many scattering processes, it is indispensable to calculate the cross sections for these processes to high accuracy, often at next-to-next-to-leading order, or, in some cases, even higher~\cite{Amoroso:2020lgh}.
The common strategy to deal with the occurring Feynman integrals in such calculations is to first express all integrals in terms of a basis of master integrals.
This so-called reduction is furthermore a key ingredient in many methods to calculate the master integrals themselves~\cite{Henn:2013pwa,Argeri:2014qva,vonManteuffel:2015gxa,Francesco:2019yqt,Hidding:2020ytt}.

In the last few years, new techniques have been explored and applied to integral reduction problems, employing
syzygy equations \cite{Gluza:2010ws,Schabinger:2011dz,Ita:2015tya,Boehm:2017wjc,Kosower:2018obg,vonManteuffel:2020vjv},
algebraic geometry \cite{Larsen:2015ped,Boehm:2018fpv,Bendle:2019csk},
intersection numbers \cite{Mastrolia:2018uzb,Frellesvig:2019kgj,Frellesvig:2019uqt,Weinzierl:2020xyy,Frellesvig:2020qot},
finite field and interpolation techniques \cite{vonManteuffel:2014ixa,Peraro:2016wsq,Smirnov:2019qkx,Klappert:2019emp,Peraro:2019svx,Klappert:2020aqs},
or special integral representations \cite{Liu:2017jxz,Liu:2018dmc,Wang:2019mnn,Guan:2019bcx}.
But unfortunately, there is still no general algorithm known that directly reduces a given set of integrals in a target-oriented manner.
Hence, in most cases the Laporta algorithm \cite{Laporta:2001dd} remains the method of choice.
Several public implementations of the algorithm exist~\cite{Anastasiou:2004vj,vonManteuffel:2012np,Smirnov:2014hma,Maierhoefer:2017hyi},
and the implementations are continuously improved to be able to handle increasingly complicated reduction problems.

In this article we present the version \code{2.0} of \kira{}, a Feynman integral reduction program based on Laporta's algorithm that was first introduced in Ref.~\cite{Maierhoefer:2017hyi}.
The most prominent new feature is the application of finite field methods to reconstruct the coefficients appearing in the integral reduction formulas with the help of \firefly{}~\cite{Klappert:2019emp,Klappert:2020aqs}.
In many cases this leads to reduced main memory usage and, depending on the underlying problem, also to reduced overall runtime.
Furthermore, the reduction can be parallelized on computer clusters using \mpi{}~\cite{mpi_forum}.
Pre-release versions of \kiraZO{} have already been successfully used in several projects~\cite{Harlander:2018zpi,Artz:2019bpr,Harlander:2020duo}.

This article is organized as follows.
In section~\ref{sect:prelimineries} we introduce some definitions used throughout in this article and briefly discuss Feynman integral reduction and finite field methods for the reconstruction of multivariate rational functions.
In section~\ref{sect:features} we describe the new features of \kiraZO{} and how to use them.
Benchmarks of the new features are presented in section~\ref{sect:benchmarks}.
Section~\ref{sect:installation} explains how to obtain, compile, and install \kira{}.
We conclude in section~\ref{sect:conclusions}.

\section{Preliminaries}
\label{sect:prelimineries}

\subsection{Feynman integral reduction}
\label{sect:feynman-integral-reduction}

The primary application of \kira{} is the reduction of Feynman integrals.
Here we introduce the notation and conventions used throughout this publication.
A general Feynman integral can be parametrized as
\begin{align}
  T(a_1,\dots,a_N)
  = \int \left(\prod\limits_{i=1}^L \mathrm{d}^d\ell_i\right)
    \frac{1}{P_1^{a_1} P_2^{a_2} \cdots P_N^{a_N}},
  \label{eq:loop-momenta-integral}
\end{align}
where $P_j=q_j^2-m_j^2$, $j=1,\dots,N$, are the inverse propagators (omitting the Feynman prescription).
The momenta $q_j$ are linear combinations of the loop momenta $\ell_i$, $i=1,\dots,L$ for an $L$-loop integral, and external momenta $p_k$, $k=1,\dots,E$ for $E+1$ external legs (or $E=0$ for vacuum integrals), and $m_j$ are the propagator masses.
The $a_j$ are the (integer) propagator powers.
The set of inverse propagators must be complete and independent in the sense that every scalar product of momenta can be uniquely expressed as a linear combination of the $P_j$, squared masses $m_j^2$, and external kinematical invariants.
The number of propagators is thus $N=\frac{L}{2}(L+2 E+1)$ including auxiliary propagators that only appear with $a_j\le 0$.

Integrals of the form \eqref{eq:loop-momenta-integral} for different values of $a_j$ are in general not independent.
\emph{Integration-by-parts} (IBP) identities \cite{Tkachov:1981wb,Chetyrkin:1981qh} and \emph{Lorentz-invariance} (LI) identities \cite{Gehrmann:1999as}, as well as symmetry relations lead to linear relations between them.
These identities can be used to express all integrals through linear combinations of master integrals, which serve as a basis.

\kira{} employs a variant of the Laporta algorithm \cite{Laporta:2001dd}:
IBP, LI, and symmetry relations are generated for different values for the $a_j$, resulting in a linear system of equations.
This system of equations is then systematically solved with a Gauss-type elimination algorithm to express integrals which are regarded more complicated in terms of simpler integrals.

To systematically classify integrals, first, integrals are assigned to so-called topologies based on their respective sets of propagators, i.e.\ their momenta and masses.
Multiple topologies are handled by assigning a unique topology ID to each topology.
Furthermore, each integral is assigned a sector
\begin{align}
  S = \sum\limits_{j=1}^N 2^{j-1}\,\theta(a_j - \tfrac{1}{2}),
  \label{eq:sector-number}
\end{align}
where $\theta(x)$ is the Heaviside step function.
A sector $S$ is called subsector of another sector $S'$ (with propagators powers $a'_j$) if $S<S'$ and $a_j\le a'_j$ for all $j=1,\dots,N$.
We denote as top-level sectors those sectors which are not subsectors of other sectors that contain Feynman integrals occurring in the reduction problem at hand.
As a measure of complexity it is useful to define the number
\begin{equation}
  t = \sum_{j=1}^N \theta(a_j - \tfrac{1}{2}),
  \label{eq:t-definition}
\end{equation}
of propagators with positive power, the sum $r$ of all positive powers, and the negative sum of all non-positive powers $s$,
\begin{equation}
  r = \sum_{j=1}^N a_j \theta(a_j - \tfrac{1}{2}),\qquad
  s = -\sum_{j=1}^N a_j\theta(\tfrac{1}{2} - a_j).
  \label{eq:rs-value}
\end{equation}
These values are used as limits in the choice of the sets $a_j$ for which the IBP, LI, and symmetry relations are generated.
The sets of $a_j$ are chosen such that $r\le r_{\max}$ and $s\le s_{\max}$, where $r_{\max}$, and $s_{\max}$ are chosen large enough to cover all relevant integrals in the reduction process, but usually no larger.

Symmetry relations either relate integrals within the same sector, or between two different sectors of the same or different topologies.
In case of a symmetry between two sectors, each integral can be expressed as a linear combination of integrals in the respective other sector.
Hence, master integrals may only appear in that sector which is considered simpler in the chosen ordering of sectors.
In case of a relation between two topologies, integrals are always mapped from the topology with higher ID to the one with lower ID, which is in that sense considered simpler.

\subsection{Finite field and interpolation techniques}
\label{sect:interpolation}

Integral reductions with the Laporta algorithm~\cite{Laporta:2001dd} for state-of-the-art problems are usually very expensive in terms of required CPU time and also main memory usage.
This is due to the huge number of equations in the system to solve, the growth of intermediate expressions while solving the system, and the size of the rational functions that appear as coefficients.
In particular, the coefficients of intermediate results are typically more complicated than those appearing in the final result.
Simplifying those coefficients by algebraic means, e.g.\ with \code{Fermat}~\cite{Fermat}, is very time consuming and memory intensive.

These problems can be mitigated by solving the system over finite fields~\cite{Kauers:2008zz}.
In practice, one uses prime fields $\mathbb{Z}_p$ with characteristic $p$, where $p$ is the defining prime.
Because of the 64-bit architecture of modern CPUs, $p$ is usually chosen to be a large 63-bit prime (so that the sum of two elements of $\mathbb{Z}_p$ still fits into 64 bits).
One can then replace all variables in the system by integers in $\mathbb{Z}_p$ and perform all operations modulo $p$.
This way, all coefficients are mapped to 64-bit integers, independent of the size of the original rational functions.
On one hand this reduces the memory needed for the coefficients, and on the other hand, arithmetic operations on $\mathbb{Z}_p$ are performed in constant time, independent of the size of the original coefficients, utilizing the CPU's native integer operations.

These methods have already been used in the first version of \kira{} to eliminate linearly dependent equations from the system~\cite{Kant:2013vta}.
This reduces the size of the system which has to be solved algebraically and already leads to significant performance improvements.
Furthermore, if a list of integrals is provided that should be reduced, the information gathered in this step can be used to select only those equations which are needed to reduce the integrals from this list.
In \code{Kira}, this procedure is implemented in the software component \code{pyRed}~\cite{Maierhoefer:2017hyi}.
Once the selection of the equations is done, the system is solved analytically, employing algebraic simplifications of the rational functions.

However, it is also possible to avoid the algebraic solution altogether and reconstruct the analytic result from the solutions over finite fields by employing interpolation and rational reconstruction techniques, see e.g.\ Ref.~\cite{deKleine:2005}, as suggested in Ref.~\cite{vonManteuffel:2014ixa} in the context of IBP reductions.\footnote{For the related strategy based on generalized unitarity, the usage of advanced techniques from computer science has been pioneered in Ref.~\cite{Peraro:2016wsq}.}
Efficient algorithms for the interpolation of polynomials and rational functions from their images in a finite field have been studied in computer science for several decades, see e.g.\ Refs.~\cite{Zippel:1979,Ben-Or:1988,Kaltofen:1988,Zippel:1990,Kaltofen:2000,Kaltofen:2003,deKleine:2005,Cuyt:2011}.
The variables of the rational function are replaced by members of the finite field and the function is evaluated at this point, i.e.\ for each tuple of values for the variables one obtains the image of the rational function at this point.
These evaluations are called \emph{probes}.
The rational function is then interpolated by processing a sufficient number of probes so that one obtains the rational function over the chosen prime field.
Note that all coefficients in the numerator and denominator polynomials are, of course, mapped to the finite field.
The rational function with rational numbers in $\mathbb{Q}$ as coefficients can be reconstructed with the help of rational reconstruction (RR) algorithms~\cite{Wang:1981,Monagan:2004}.
Based on the image of the rational number in $\mathbb{Z}_p$ and the prime number $p$ of the field, these algorithms can ``guess'' the rational number in $\mathbb{Q}$.
They only succeed if both numerator and denominator of the rational number are significantly smaller than $p$.
However, this guess is unique if they succeed.
The limit on the size can be circumvented by combining the images over several prime fields with the help of the Chinese remainder theorem (CRT)~\cite{von_zur_Gathen}.
It combines the images of a rational number over two coprime numbers to a new image over the product of both coprimes.
Thus, the upper limit in the RR is increased.
It is important to note that most of the algorithms are probabilistic, i.e.\ there is a chance that they fail or provide a wrong result.
Most of the failures are triggered by hitting accidental zeros with a probability based on the DeMillo-Lipton-Schwartz–Zippel lemma~\cite{DeMillo:1978,Zippel:1979,Schwartz:1980}.
The probability of obtaining a wrong result can be reduced by performing additional checks after the termination of the algorithms.

The general strategy can be summarized as follows.
The system of equations is repeatedly solved over a prime field.
Each solution results in a probe for each master integral coefficient.
These are then handed to a rational function interpolation algorithm.
This procedure is repeated until all rational functions have been interpolated over the prime field.
Their coefficients are then passed to a RR algorithm to obtain the rational functions with coefficients in $\mathbb{Q}$.
If the RR did not succeed, the same process is repeated over additional distinct prime fields, and the results are combined with the help of the CRT until the RR succeeds.
\code{FIRE6} was the first public program implementing this strategy~\cite{Smirnov:2019qkx}.
However, it is currently limited to problems with $d$ and two scales, of which one has to be set to one.
It uses a factorization strategy which still has to be generalized to more scales.

While the simplification of rational functions can be performed in parallel to some degree, limited by their interdependencies, the evaluations of probes on a finite field are completely independent, opening the possibility for massive parallelization on many CPU cores and even nodes of a computer cluster.

Last year, two general purpose \code{C++} libraries implementing both interpolation and RR algorithms have been published, namely \code{FireFly}~\cite{Klappert:2019emp,Klappert:2020aqs}, which we chose to use in \kira{}, and \code{FiniteFlow}~\cite{Peraro:2019svx}.
\code{FireFly} requests \code{pyRed} to solve the system of equations repeatedly over the finite fields for different tuples of values of the variables and then processes the resulting probes until the master integral coefficients are successfully reconstructed over $\mathbb{Q}$.
The input system is the same system as for the algebraic reduction.
Particularly, it is already trimmed of the linearly dependent equations and of the equations which are not relevant for the selected integrals.
Exactly as in the algebraic reduction, the system is trimmed again after the forward elimination, i.e.\ the equations which are no longer relevant are dropped.
However, in contrast to the algebraic reduction, we only select the relevant master integral coefficients after the back substitution as suggested in Ref.~\cite{Klappert:2019emp}.
Hence, the interpolation of irrelevant (but potentially difficult) rational functions is avoided.
This selection does not offer any advantage in the algebraic reduction because all intermediate coefficients need to be known in order to calculate the final result.
In Sect.~\ref{sect:benchmarks}, we show some benchmarks of this newly implemented approach.

Interestingly, the forward elimination is usually the dominant part of a reduction over a finite field, whereas in an algebraic solution it is usually the other way round.
Hence, we implemented a second strategy which first performs the forward elimination using algebraic simplifications for the coefficients, and then performs the back substitution over the finite field, reconstructing the result with \firefly{} as suggested in Ref.~\cite{Klappert:2019emp}.
In Sect.~\ref{sect:schmuckstueck}, we present an example which heavily profits from this strategy.

\section{New Features in Kira 2.0}
\label{sect:features}

The following usage instructions extend those from the original \kira{} publication \cite{Maierhoefer:2017hyi} if not stated otherwise.
Note that before starting a new reduction, the temporary files and directories \code{tmp}, \code{firefly\_saves}, and \code{ff\_save} must be removed from the project directory.
If the topology definition changed, also \code{sectormappings} and \code{results} must be removed.

\subsection{Finite field reduction}
\label{sect:finite_field}

The central new feature of using the interpolation and reconstruction techniques described in Sect.~\ref{sect:interpolation} can be enabled with the option \code{run\_firefly} in the job file.
The option comes in two variants:
\code{run\_firefly:\;true} imports the entire system of equations (stored in the files \code{tmp/[topology]/SYSTEM\_*}) and performs the full reduction.
\code{run\_firefly:\;back} on the other hand performs just the back substitution.
Hence, it requires the triangular system calculated by \code{run\_triangular} (stored in the files \code{tmp/[topology]/VER\_*}) in the same or a previous run.

Per default, \firefly{} performs the factor scan (see Ref.~\cite{Klappert:2020aqs} for details) for reductions with three or more variables.
The default behaviour can be overwritten with the job-file option \code{factor\_scan:\ <true|false>}.

\firefly{} offers the possibility to calculate several probes at once by combining several coefficients for different values of the variables in a coefficient array.
This reduces the overhead due to traversing the system during its solution for the price of moderately increased main memory usage.
We refer to Ref.~\cite{Klappert:2020aqs} for more details on the implementation.
This behaviour can be enabled by the new command line option \code{-{}-bunch\_size=n}, which sets \firefly{}'s maximum bunch size to \code{n}.
It is advisable to first use the available memory to utilize as many cores as possible before spending it for an increased bunch size.

Lastly, \code{run\_firefly} supports additional parallelization with \code{MPI}~\cite{mpi_forum}, where the additional nodes are used as workers to compute the probes required by the main node which carries out the interpolation.
Each process uses its own thread pool for multithreading.
See Ref.~\cite{Klappert:2020aqs} for more details.
This feature is automatically enabled when \firefly{} is installed with \code{MPI} support enabled (see Sect.~\ref{sect:installation}).
Usually, \kira{} can then be invoked with \code{MPI} by
\begin{verbatim}
  mpiexec -n <n> [other MPI options] kira <job_file> [kira options]
\end{verbatim}
where \verb|<n>| is the number of processes to be started and \verb|<job_file>| is a usual \kira{} job file with \firefly{} enabled.
The details of course depend on the \code{MPI} implementation available on the system.
Since the different processes have to communicate with each other, the best performance is obtained by assigning all cores of a machine to a single process of \kira{} with the \code{-{}-parallel} command-line option.
\mpi{} should only be used for multiple machines.

The benchmarks in Sect.~\ref{sect:benchmarks} show how the different variants of the option \code{run\_firefly} behave.

Especially on systems with older versions of \code{glibc}\footnote{Notable examples are most computer clusters with Intel CPUs.} we strongly recommend to use a library like \code{jemalloc}~\cite{jemalloc} that replaces the \code{malloc} function by an implementation that is optimized for high performance memory allocation under multithreaded workloads (see Sect.~\ref{sect:installation}).

\subsection{User-defined systems of equations}
\label{subsect:user-defined-system}

Since \kira{}\;\code{1.2} it has been possible to use \kira{} to solve a system
of linear equations provided by the user.
This functionality has been significantly improved and extended.
The option to provide the system of equations reads
\begin{verbatim}
input_system: {files: [<file1>,<file2>,...], otf: <true|false>,
               size: <n>, config: <true|false>}
\end{verbatim}
where the specification of \code{size}, \code{otf}, and \code{config} is optional.
\begin{itemize}
  \item It is now possible to provide several files at once.
    Files may optionally be compressed with \code{gzip}.
    It is also possible to pass the names of directories, in which case all
    regular files with the file extension \code{.kira} resp.\ \code{.kira.gz}
    within these directories are used.
  \item With \code{otf:\;true} (default: \code{false}), the
    ``on-the-fly solver'' is used, i.e.\ each equation is passed to the solver
    immediately after it is parsed, yielding reduced main memory usage
    and slightly reduced runtime.
    If this option is used, it is crucial that the equations in the files are
    approximately ordered by complexity, starting with simpler equations, where
    simpler refers to the most complicated integral in the equation.
    The system doesn't have to be strictly ordered, but equations of similar
    complexity should be close to each other, otherwise the runtime may increase
    drastically.
    Files are read in the order in which they are passed to the \code{files}
    option.
    In case of a directory, the files in the directory are ordered
    lexicographically by their names.
  \item If the option \code{otf} is used, the option \code{size} can be used
    to give the total number of equations.
    If \code{size} is omitted, the files will be read once in advance to
    determine the number of equations, which costs some extra time,
    and a second time to solve them.
  \item If the option \code{config:\;true} is set (default: \code{true}),
    the topology definitions from the \code{config} directory will be used.
    Otherwise not.
    If \code{config:\;false}, the variables occurring in the coefficients are
    determined automatically.
    Besides the integral notation with a topology name and a list of integer
    indices, it is now possible to directly use 64-bit integral weights
    (described below).
    This circumvents the limitation that, at least in the case of a
    single-indexed topology, the number of integrals per topology is limited
    to $2^{32}$.
\end{itemize}
Of course, user-defined systems can be used together with \firefly{}.
Note that the old format for the option \code{input\_system}, where only a single file name can be given, still works, but is considered deprecated.

\medskip\pagebreak[2]
\noindent
\textit{User-defined weights}

Internally, \kira{} represents every integral by a 64-bit unsigned integer.
This ``integral weight'' serves as a measure of complexity of an integral.
The higher the weight, the more complicated the integral in the chosen
ordering.
When dealing with user-provided systems of equations, it is now possible
for the user to choose the weights of the objects in the equations manually.
This simply works by using 64-bit unsigned integers everywhere, where
previously only integrals with a topology name and indices (like
e.g.\ \code{T[2,1,1]}) could be used.
An equation file with user-defined weights may look as in the following example:
\begin{verbatim}
# three equations with user-defined weights
18084767254708224*(-4)
18014398510530560*(d)

19140298417373185*(-2+d)
18014398510530560*(-4)

20266198324215809*(-4+d)
19140298417373185*(-4)
\end{verbatim}
This file contains three equations, each of length two, separated by empty lines.
It is implied that the expression for each equation equals zero.
User-defined weights may have values from 0 to $2^{64}-2$.
The weight $2^{64}-1$ is reserved for internal purposes.
Each term, i.e.\ a 64-bit unsigned integer and the coefficient separated by a character $\ast$, has to occupy a separate line.
Furthermore no spaces are allowed.
All equations must be separated by empty lines.

Note that the option \code{preferred\_masters} cannot be used together with
user-defined weights.
However, it is straight forward to choose a preferred basis manually by
assigning sufficiently low weights to the preferred masters.
Linear combinations as basis elements or entire amplitudes can be easily
handled by adding a corresponding equation.
Note that the concept of sectors is not defined for user-defined weights.
I.e.\ that options like sectorwise forward elimination or sectorwise
iterative reduction (see Sect.~\ref{subsect:iterativ reduction}) cannot be used.

\subsection{Iterative reduction}
\label{subsect:iterativ reduction}

In \kira{}\;\code{1.1} the option \code{select\_masters\_reduction} \cite{kira:1_1} was introduced to calculate only the coefficients of a subset of master integrals, effectively setting all other master integrals to zero.
By setting certain integrals to zero, the size of the equations and hence the main memory consumption is reduced.
On the other hand, to achieve the full reduction, the procedure must be repeated, possibly in parallel on several machines, until the coefficients of all master integrals are known.
However, the usage of this option is quite cumbersome.
The set of selected master integrals for each node has to be provided in the corresponding job file for the node, and the databases with the partial reductions have to be merged manually at the end~\cite{Maierhofer:2018gpa}.

With the option \code{iterative\_reduction}, a similar and fully automated strategy for the iteration over the master integrals is now available.
The option comes in two variants:
\begin{verbatim}
iterative_reduction: masterwise
\end{verbatim}
performs separate reductions for each master integral sequentially and
\begin{verbatim}
iterative_reduction: sectorwise
\end{verbatim}
performs separate reductions for all master integrals in each sector sequentially.
Iterative reduction can be used both with \code{run\_back\_substitution} (i.e.\ \code{Fermat}) and \code{run\_firefly}.
If \code{MPI} is used, the parallelization across several nodes is done at the level of each iteration step, i.e.\ one master integral resp.\ sector at a time.
\code{run\_back\_substitution} will create separate databases for the partial reductions, i.e.\ per master integral or per sector.
These databases will be automatically merged into a single database with the full result once all partial reductions are complete.

In combination with \firefly{}, another interesting effect comes into play.
From experience we know that the complexity of the coefficients in the reduction varies strongly between different master integrals.
Hence, for a master integral that comes with simple coefficients, fewer probes (often by orders of magnitude) have to be calculated than for a master integral with complicated coefficients.
Moreover, equations that reduce integrals to zero, because all master integrals appearing in the reduction are set to zero, can be removed from the system in the respective iteration step.
While these effects increase the performance of the reduction, the increased overhead due to solving the system many more times has the opposite effect.

The primary use case for this feature is to decrease the required amount of main memory.
It is difficult to predict the effect on the performance.
In Sect.~\ref{sect:iterative_bench}, we show an example where the memory usage is reduced significantly, but the overall runtime increases.
Usually, \code{sectorwise} should be preferred over \code{masterwise} unless the further reduced memory consumption is crucial for being able to run the reduction at all on the available machines.

\subsection{Master equations}
\label{subsect:master-equations}

In some cases it can be useful to treat linear combinations of integrals as basis elements, i.e.\ as if they were master integrals.
The most prominent example occurs in the context of systems of coupled differential equations for master integrals, where a convenient basis choice can lead to a particularly simple form of the differential equations~\cite{Henn:2013pwa,Argeri:2014qva}.

Effectively, \kira{} handles such linear combinations by introducing a new integral-like object with a small weight (in the sense of the integral ordering) that serves as a master integral, and adding an equation to the system that equates this object with the given linear combination.
Hence, we refer to these equations as \emph{master equations}.
Master equations are defined similarly to preferred master integrals, using the same input file.
The following example illustrates the syntax to define linear combinations:
\begin{verbatim}
# A master equation with 3 elements
box[1,1,1,1]*(1)
box[1,1,1,2]*(s)
box[1,2,1,1]*(s)

# A master equation with 1 element
box[0,1,0,1]*(s)

# Two individual preferred masters. No empty line needed in between,
# i.e. the notation from Kira 1.1 is still accepted.
box[1,0,1,0]
box[1,0,0,2]
\end{verbatim}
Note that the factor \verb|1| in the first master equation is required to indicate that the integral \verb|box[1,1,1,1]| is part of a linear combination.
Like with usual preferred master integrals, the file with the chosen basis is passed to \kira{} in the job file with the option \code{preferred\_masters:\;"basisFile"}, where \code{basisFile} is the name of the file in which the basis is defined.

All master equations are enumerated starting with $1$ and represented as \code{BASISLC[n]}, where \code{n} is the number of the master equation in the order in which they are defined in the preferred basis file.

\subsection{Sectors in big-endian binary notation}
\label{subsect:reverse-bit-notation}

\kira{} supports sectors in the big-endian binary notation both in the job file and in the \code{integralfamilies.yaml} file.
One can just replace the sector in the previous notation by the big-endian binary notation, e.g.\
\begin{verbatim}
top_level_sectors: [b111111100]
\end{verbatim}
The first letter \texttt{b} is important.
It tells \kira{}'s parser that the following numbers belong to the big-endian binary notation.
In the example above the big-endian binary notation is defined for integrals with 9 propagators. The first 7 propagators are in the denominator and the last two are the irreducible scalar products in the numerator.

\subsection{Setting all integrals in a sector to zero}
\label{subsect:zero_sectors}

With the option \code{zero\_sectors} it is possible to set entire sectors, i.e.\ all integrals belonging to these sectors, to zero.
All sectors passed to this option will be added to the automatically determined list of trivial sectors.
The option can be set per topology in the file \code{integralfamilies.yaml}, e.g.\ \code{zero\_sectors:\;[b111111100]} to set sector $127$ of the respective topology to zero.

\subsection{Export reduction rules with kira2file}
\label{subsect:kira2file}

\kira{} can export results in formats readable by \code{Mathematica} and \code{FORM} with the options \code{kira2math} and \code{kira2form}.
Additionally, with \code{kira2file} the results can now be exported in the format that is compatible with the input format of user-defined systems of equations (see Sect.~\ref{subsect:user-defined-system}).
I.e.\ files exported in this format can be used as input in further \kira{} runs.

\subsection{Factor out prefactors with FireFly}
\label{subsect:factor-prefactors}

Recently, algorithms and tools have been published that make it possible to determine the denominators of all coefficients in the result of the reduction without performing a full reduction \cite{BSMF_1992__120_3_371_0,Smirnov:2020quc,Usovitsch:2020jrk}.
When the denominators are known, they can be divided out of the coefficients, so that only the numerators, i.e.\ polynomials instead of rational functions, remain.
This not only reduces the number of terms which have to be interpolated by a factor of roughly two (assuming that numerator and denominator are of similar complexity), but moreover simplifies the interpolation, because polynomial interpolation algorithms can be used instead of the much more involved algorithms for rational functions.\footnote{\code{FireFly} still performs the interpolation of a rational function for technical reasons. However, the additional runtime to identify trivial denominators is marginal for multiscale problems.}
Of course, the interpolation also simplifies if numerators or partial factors are known and divided out.

With the option \code{insert\_prefactors}, \kira{} can load a list of prefactors when \code{run\_firefly} is used.
During the interpolation with \code{FireFly}, the results of the reductions over finite fields are divided by these prefactors.
In the final result, all factors are restored.
Each prefactor has to be assigned to the integral in whose reduction it appears and the master integral from whose coefficient it is divided out.
This is done in a file, where in one line the integral to be reduced is listed, followed by a factor that must always be one.
In the following lines, the master integrals are listed, one per line, followed the factors to be divided out of their coefficients.
Integrals and factors are separated by a multiplication symbol ``\code{*}''.
Example:
\begin{verbatim}
doublebox[1,1,1,1,1,1,1,1,-2,0,0] * 1
doublebox[0,0,1,0,0,1,1,0,0,0,0] * 1/((d-8)*(d-6)^3*(d-5)^3*(d-4)^3*...
doublebox[0,0,1,1,0,0,1,0,0,0,0] * 1/((d-8)*(d-6)^3*(d-5)^3*(d-4)^3*...
doublebox[0,0,1,1,0,1,1,1,0,0,0] * 1/((d-8)*(d-6)^2*(d-5)^2*(2*d-11)*...
...
\end{verbatim}
Here, in the reduction of the integral \verb|doublebox[1,1,1,1,1,1,1,1,-2,0,0]|,
the coefficient $1/((d-8)(d-6)^3(d-5)^3(d-4)^3\dots$ will be divided out of the
coefficients of the master integrals \verb|doublebox[0,0,1,0,0,1,1,0,0,0,0]|
and \verb|doublebox[0,0,1,1,0,0,1,0,0,0,0]|,
and $1/((d-8)(d-6)^2(d-5)^2(2d-11)\dots$ will be divided out of the
coefficient of the integral \verb|doublebox[0,0,1,1,0,1,1,1,0,0,0]|.
Several such prescriptions for different integrals to be reduced can be provided in a single file, separated by empty lines.
As the notation suggests, it is also possible to use rational functions as prefactors.
The file name is simply passed to the option \code{insert\_prefactors} in the job file.
The above example is taken from \code{examples/insert\_prefactors} with the prefactors listed in the file \code{xints}.

Note that usually such a factorization only works in a specific basis of master integrals.
This basis must then by chosen with the option \code{preferred\_masters}.
In the above example, the proper basis is chosen in the file \texttt{preferred}.

\subsection{General propagators}
\label{subsect:general-propagators}

Some methods for the computation of Feynman integrals require more general propagators than the usual form $q_j^2-m_j^2$~\cite{Papadopoulos:2014hla,Hidding:2017jkk,Papadopoulos:2019iam}.
\kira{} now offers more freedom to define the propagators.
For example, one can define the propagator $(x (q_1^2 - m_1^2) + (1-x) (q_2^2 - m_2^2))^2$ obtained by the combination of two propagators into a one-dimensional Feynman-parameter integral according to
\begin{equation}
  \frac{1}{(q_1^2 - m_1^2) (q_2^2 - m_2^2)}
  = \int\limits_0^1\mathrm{d}x\;\frac{1}{(x (q_1^2 - m_1^2) + (1-x) (q_2^2 - m_2^2))^2}.
\end{equation}
As the following example shows, defining general propagators is straight forward:
\begin{verbatim}
integralfamilies:
  - name: "box"
    loop_momenta: [k1]
    top_level_sectors: [b1110]
    propagators:
      - [ "k1^2 + 2*k1*p1 + p1^2 + (-m2 - 2*k1*p1 - p1^2)*x", 0 ]
      - [ "(k1+p1+p2)^2-m2", 0 ]
      - [ "(k1+p1+p2+p3)^2", 0 ]
      - [ "k1^2", 0 ]
\end{verbatim}
The full example can be found in \code{examples/general\_propagators}.
Here, all four propagators are defined using the notation for general propagators.
However, whenever possible, we recommend to use the standard notation with the unsquared propagators momentum.
The reason is that \kira{} is currently not able to apply symmetry relations to integrals with general propagators in the numerator (i.e.\ with negative powers).

For compatibility with the format used in \code{Reduze\;2}~\cite{vonManteuffel:2012np}, it is also possible to define a propagator as
\begin{verbatim}
- { bilinear: [ [ "l3", "q1" ], 0 ] }
\end{verbatim}
meaning that the propagator has the form $l_3\cdot q_1$ (see example \code{example/aah-nl-sing}).

In particular, these notations can be used to define propagators in which loop momenta appear only linearly, e.g.\ in HQET (see e.g.\ \cite{Grozin:2004yc}).
The different notations can be mixed in the same topology definition.

\subsection{Add sectors for preferred master integrals}
\label{subsect:add-missing-sectors}

\kira{} now ensures that for each integral listed in \code{preferred\_masters}, a system of equations will be generated in the sector to which this integral belongs, even if the sector is not requested for reduction in the job file.
The limits for $r$ and $s$ (see Eqs.~\eqref{eq:rs-value}) are chosen based on values from the job file.
If only a single topology is reduced, the effect is the same as if the sector is requested for reduction with the respective values of $r$ and $s$.
But if several topologies are reduced, this is the only way to include this sector together with the symmetry related sectors of lower topologies that are generated to map the master integrals across topologies.

One possible application of this feature is to find magic relations, i.e.\ relations originating from higher sectors that reduce the number of master integrals~\cite{Frellesvig:2018ymi}, in lower topologies w.r.t.\ the currently reduced topology.
An example to illustrate the effect in case of a single topology can be found in \code{examples/magic\_relations}.

\subsection{Generate a system of equations for later reduction as user-defined system}
\label{sect:generate_input}

The option \code{generate\_input:\;\{level:\;<n>\}} with \code{<n>}=\code{0} generates the system of equations for the given seeds, selects a linearly independent subsystem and writes the equations into files in the directory \code{input\_kira}.
The integrals are represented as integer weights, and the generated files are suitable to be read with the option \code{input\_system} using the on-the-fly solver (see Sect.~\ref{subsect:user-defined-system}).
If \code{<n>} is an integer $\ge 1$, a system of equations will be generated for each subsector of the chosen top-level sector with \code{<n>} lines less,
and one for the remaining sectors.
One possible application is to reduce the amount of memory (often by more than $50\,\%$) that is needed to generate the entire system and for the selection of linearly independent equations.
The generation time on the other hand will be significantly longer, though.
Note that at this point it is not possible to select subsystems of equations to solve specific integrals (e.g.\ with \code{select\_mandatory\_list}).
However, this can be done in a further run, where the generated system is solved as a user-defined system.

The option \code{amplitude\_translate} reads a linear combination of integrals from a file and assigns an integer weight to it, representing the expression (usually an amplitude or a part of it).
Multiple amplitudes, separated by empty lines can be provided in the same file.
All integrals are converted into weights and the result is written into a file in the directory \code{input\_kira} in the form of an equation that can be added to a system of equations that is solved with the option \code{input\_system}.
The weight of the object representing the expression will be $2^{64}-1$ for the first expression, counting downwards if further expressions are defined.
These weights can be used to refer to the expressions e.g.\ with \code{select\_mandatory\_list}.

An example is provided in \code{examples/aah-nl-sing}.

%
%

\section{Benchmarks}
\label{sect:benchmarks}

In this section we present benchmarks for the new features in \kiraZO{}.
The numbers, especially the runtime, should be read with uncertainties of a few percent in mind.
These are caused by using different nodes on a computer cluster, which are nominally equal but of course behave slightly differently due to a difference in the quality of the CPUs and thermal effects.
Moreover, the IO operations are performed on the cluster filesystem with fluctuating performance depending on the overall workload on the entire cluster.

\subsection{Runtime reduction with bunches and MPI}

\begin{figure}[ht]
  \centering
  \includegraphics{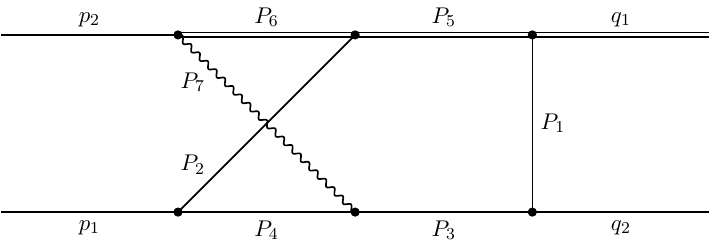}
  \caption{The non-planar double box \code{topo5} which occurs, e.g., in virtual corrections to single top production at NNLO.}
  \label{fig:topo5}
\end{figure}
To illustrate the impact of the overhead reduction with the \code{-{}-bunch\_size} command-line option and the scaling with several nodes utilizing \code{MPI}, we chose the IBP reduction of the topology \code{topo5} shown in Fig.~\ref{fig:topo5} with the propagators
\begin{equation}
  \begin{gathered}
    P_1 = k_1^2,\quad
    P_2 = k_2^2,\quad
    P_3 = (q_2-k_1)^2,\quad
    P_4 = (p_1-k_2)^2,\\
    P_5 = (q_1+k_1)^2 - m_1^2,\quad
    P_6 = (q_1+k_1-k_2)^2 - m_1^2,\\
    P_7 = (q_1-p_2+k_1-k_2)^2 - m_2^2,\quad
    P_8 = (k_1-p_1)^2,\quad
    P_9 = (k_2-q_2-p_2)^2,
  \end{gathered}
  \label{eq:topo5_prop}
\end{equation}
where $P_8$ and $P_9$ are auxiliary propagators.
The scalar products of the external momenta can be expressed through the kinematical invariants by
\begin{equation}
  \begin{gathered}
    p_1^2 = p_2^2 = q_2^2 = 0, \quad q_1^2 = m_1^2, \quad (p_1+p_2)^2 = s,\\
    (q_2-p_1)^2 = t, \quad (q_2-p_2)^2 = m_1^2 - s - t.
  \end{gathered}
\end{equation}
This example can be found in \code{examples/topo5}.
We perform the reduction with $r = 7$ and $s = 4$, which is sufficient for the virtual NNLO corrections to the amplitude of single top production.
This results in about 1.6\,million probes in total over four prime fields to complete the reduction.
The runtime of the probes is dominated by the forward elimination with a contribution of 97\,\%.
The calculations were performed on cluster nodes equipped with two Intel Xeon Platinum 8160 processors with 24 cores each and 192\,GiB of RAM in total with hyper-threading disabled.
On each node, we utilize all cores with the command line option \code{--{}--parallel=48}.
\code{topo5} has also been studied in Ref.~\cite{Klappert:2020aqs} with emphasis on \firefly{}'s options.

In Tab.~\ref{tab:topo5_bunches}, we vary the maximal bunch size of \firefly{} with the command-line option \code{-{}-bunch\_size} as described in Sect.~\ref{sect:finite_field}.
\begin{table}[ht]
  \begin{center}
    \caption{Reduction of \code{topo5} with $r=7$, $s=4$ utilizing \firefly{}, i.e.\ \code{run\_firefly:\ true}, and the command-line option \code{-{}-bunch\_size}.
    As a comparison we present the algebraic reduction with \kira{} and \fermat{}, i.e.\ using \code{run\_triangular:\ true} and \code{run\_back\_substitution:\ true}.}
    \label{tab:topo5_bunches}
    {\renewcommand{\arraystretch}{1.3}
    \begin{tabular}{c|c c c c}
      \toprule
      \code{-{}-bunch\_size=} & Runtime & Memory & \begin{tabular}{@{}c@{}}CPU time \\ per probe\end{tabular} & \begin{tabular}{@{}c@{}}CPU time \\ for probes\end{tabular} \\
      \midrule
      1 & 18\,h & 40\,GiB & 1.73\,s & 95\,\% \\
      2 & 14\,h & 41\,GiB & 1.30\,s & 94\,\% \\
      4 & 11\,h & 46\,GiB & 1.00\,s & 93\,\% \\
      8 & 10\,h 15\,min & 51\,GiB & 0.91\,s & 92\,\% \\
      16 & 9\,h 45\,min & 63\,GiB & 0.85\,s & 92\,\% \\
      32 & 9\,h 30\,min & 82\,GiB & 0.84\,s & 92\,\% \\
      64 & 9\,h 30\,min & 116\,GiB & 0.83\,s & 92\,\% \\
      \midrule
      \kira{} $\oplus$ \fermat{} & 82\,h & 147\,GiB & - & - \\
      \bottomrule
    \end{tabular}}
  \end{center}
\end{table}
In general, the runtime decreases at the cost of additional memory, as expected.
Increasing the bunch size from 1 to 2 costs about 1\,GiB of additional RAM but already decreases the runtime by roughly 25\,\% by speeding up the average time to solve the system for one data point with \pyred{} by a similar percentage.
The step to a bunch size of 4 again yields a runtime decrease of 20\,\% but already costs 5\,GiB.
In the following steps, the gain in runtime becomes smaller while the amount of additional memory required increases faster.
Note that this example is mainly limited by the cost of computing probes with \pyred{} as shown in the last column, i.e.\ the internal calculations in \firefly{} only play a minor role.
The share of the probes at the total CPU time for the reduction only decreases marginally.

In Tab.~\ref{tab:topo5_mpi}, we show the behaviour of the reduction of \code{topo5} when using multiple nodes with \code{Intel\textsuperscript{\textregistered} MPI}~\cite{intelmpi}.
\begin{table}[ht]
  \begin{center}
    \caption{Reduction of \code{topo5} with $r = 7$ and $s = 4$ utilizing \firefly{}, i.e.\ \code{run\_firefly:\ true}, and multiple nodes with \code{MPI}.
    The speed-up is measured with respect to the reduction with bunch size 1 in Tab.~\ref{tab:topo5_bunches}.
    As a comparison we show the algebraic reduction with \kira{} and \fermat{} using \code{run\_triangular:\ true} and \code{run\_back\_substitution:\ true}.}
    \label{tab:topo5_mpi}
    {\renewcommand{\arraystretch}{1.3}
    \begin{tabular}{c|c c c}
      \toprule
      \# nodes & Runtime & Speed-up & CPU efficiency \\
      \midrule
      1 & 18\,h & 1.0 & 95\,\% \\
      2 & 10\,h 15\,min & 1.8 & 87\,\% \\
      3 & 7\,h 15\,min & 2.5 & 82\,\% \\
      4 & 5\,h 45\,min & 3.1 & 76\,\% \\
      5 & 5\,h 30\,min & 3.3 & 65\,\% \\
      \midrule
      \kira{} $\oplus$ \fermat{} & 82\,h & - & - \\
      \bottomrule
    \end{tabular}}
  \end{center}
\end{table}
The additional nodes are used as pure workers solving the IBP system with \pyred{} as described in Sect.~\ref{sect:finite_field}.
They require about 9\,GiB of RAM.
Doubling the number of cores yields a speed-up of 1.8.
Using three nodes increases the speed-up to 2.5 and four to 3.1.
Going to five nodes, the speed-up only marginally increases further to 3.3.
Therefore, using even more nodes does not seem to be worthwhile for this example.
However, the calculation is still limited by computation of the probes since the percentage of the CPU time for the probes does not vary much.
On the other hand, the CPU efficiency drastically decreases, i.e.\ some of the cores are idle for some time during the calculation.
This is due to the algorithms implemented in \firefly{}, which cannot process arbitrary probes.
Instead, new probes are scheduled based on intermediate results.
This is especially relevant for the first prime field, where the structure of the functions is not known yet.
For more details, we refer to Refs.~\cite{Klappert:2019emp,Klappert:2020aqs}.

Thus, both features can be used to significantly decrease the runtime of reductions which are limited by the evaluations of the probes.
Bunches should be used if there is unused memory on the system, \mpi{} if there are more computers or cluster nodes available.
As already mentioned in Sect.~\ref{sect:finite_field}, one should not use \mpi{} when only using a single machine, mainly because one thread is reserved solely for communication.
Of course, both features can also be combined if enough resources are available.
It might also be worthwhile to monitor a long and difficult calculation and adapt the settings to the current status, e.g.\ by increasing the number of nodes when the prime field changes, because all probes are queued in the beginning, and decreasing the number when \firefly{} only interpolates the coefficients.
In Sect.~\ref{sect:double_pentagon} we show an example which is mainly limited by the interpolation with \firefly{} and, thus, the potential speed-up by increasing the bunch size or by using MPI is relatively small, because both features effectively reduce the wall clock time to compute the probes.

\subsection{Reducing the memory footprint with iterative reduction}
\label{sect:iterative_bench}

The iterative reduction described in Sect.~\ref{subsect:iterativ reduction} can be used to reduce the memory footprint by setting master integrals to zero, either all except one master integral (``masterwise'') or all master integrals except those in one sector at a time (``sectorwise'').
We again study \code{topo5} as an example on the same machines with two Intel Xeon Platinum 8160 processors with 24 cores each and 192\,GiB of RAM in total with hyper-threading disabled.
As shown in Tab.~\ref{tab:topo5_iterative}, the required memory reduces by more than a factor of 4 for the reduction with \firefly{} when using the sectorwise iterative reduction.
\begin{table}[ht]
  \begin{center}
    \caption{Sectorwise iterative reduction of \code{topo5} with $r = 7$ and $s = 4$
      utilizing \firefly{}, i.e.\ \code{run\_firefly:\ true}, and \code{-{}-parallel=48}.}
    \label{tab:topo5_iterative}
    {\renewcommand{\arraystretch}{1.3}
    \begin{tabular}{c c|c c}
      \toprule
      Mode & Iterative & Runtime & Memory \\
      \midrule
      \multirow{2}{*}{\kira{} $\oplus$ \firefly{}} & - & 18\,h & 40\,GiB \\
      & sectorwise & 33\,h 15\,min & 9\,GiB \\
      \bottomrule
    \end{tabular}}
  \end{center}
\end{table}
However, the runtime increases by 80\,\% in this example.

\subsection{Combining algebraic forward elimination with finite field reduction}
\label{sect:schmuckstueck}

Our next benchmark is an example from the study of conformal integrals in position space with the propagators\footnote{
Many thanks to Raul João Pereira for this example.}
\begin{equation}
  \begin{gathered}
    P_1 = k_1^2,\quad
    P_2 = k_2^2,\quad
    P_3 = k_3^2,\quad
    P_4 = (p_1-k_1)^2,\quad
    P_5 = (p_1-k_2)^2,\\
    P_6 = (p_1-k_3)^2,\quad
    P_7 = (p_2-k_1)^2,\quad
    P_8 = (p_2-k_2)^2,\quad
    P_9 = (p_2-k_3)^2,\\
    P_{10} = (k_1-k_2)^2,\quad
    P_{11} = (k_1-k_3)^2,\quad
    P_{12} = (k_2-k_3)^2.
  \end{gathered}
  \label{eq:topoRaul_prop}
\end{equation}
The top-level sector is $4095$, i.e.\ all propagators may appear with positive powers.
The scalar products of the external momenta can be expressed through new variables $z$ and~$z_b$:
\begin{equation}
  p_1^2=z z_b , \quad p_2^2 = 1, \quad p_1 p_2 = (1-z)(1-z_b).
\end{equation}
We chose $r = 17$ and $s = 0$ for the benchmark.

The reductions are performed on a machine with two Intel Xeon Gold 6138 with 20 cores each and 768\,GiB of RAM in total with hyper-threading enabled.
All cores including hyper-threading are utilised with the option \code{-{}-parallel=80}.
Tab.~\ref{tab:schmuckstueck} compares different strategies for the reduction.
\begin{table}[ht]
  \begin{center}
    \caption{Different strategies for the reduction of the topology defined by the propagators in Eq.~\eqref{eq:topoRaul_prop}.
    The reduction specific values are $r = 17$ and $s = 0$.
    The option \code{run\_initiate} is used to initialise the system for all strategies, but is expensive enough to warrant a separate entry in the table.}
    \label{tab:schmuckstueck}
    {\renewcommand{\arraystretch}{1.3}
    \begin{tabular}{c|c c c c c}
      \toprule
      Mode & Runtime & Memory & Probes & \begin{tabular}{@{}c@{}}CPU time \\ per probe\end{tabular} & \begin{tabular}{@{}c@{}}CPU time \\ for probes\end{tabular} \\
      \midrule
      \code{run\_initiate} & 5\,h 20\,min & 128\,GiB & - & - & - \\
      \midrule
      \begin{tabular}{@{}c@{}}\code{run\_triangular} + \\ \code{run\_back\_substitution}\end{tabular} & >\,14\,d & \textasciitilde\,540\,GB & - & - & - \\
      \midrule
      \code{run\_firefly:\ true} & 6\,d 3\,h & 670\,GiB & 108500 & 370\,s & 100\,\% \\
      \midrule
      \begin{tabular}{@{}c@{}}\code{run\_triangular:} \\ \code{sectorwise}\end{tabular} & 36\,min & 4\,GiB & - & - & - \\
      \code{run\_firefly:\ back} & 4\,h 54\,min & 35\,GiB & 108500 & 12.2\,s & 100\,\% \\
      \bottomrule
    \end{tabular}}
  \end{center}
\end{table}
The initialization is already quite expensive, especially in terms of memory.
The reduction with \firefly{} is completely limited by the calculation of the probes.
Even though the total number is rather small compared to other problems, the 370\,s for each probe are extremely expensive.
98\,\% of this time is spent on the forward elimination.
However, the forward elimination can be performed in just 36\,min by the algebraic mode of \kira{}.
Using this result as starting point for \firefly{} significantly speeds up the probes to 12\,s and, therefore, the whole reduction.
Moreover, the memory footprint of the reduction improves significantly because the forward-solved system only consists of 572313 equations with 6144971 terms instead of 8922459 equations with 64009470 terms.
However, the number of distinct coefficients increases from 205 to 983420 more complicated ones, because the forward elimination already partially solved the system.
Thus, the time to evaluate these coefficients for each probe increases from 0.002\,s to 3.3\,s.
Moreover, the time for the back substitution increases from 6.7\,s to 8.9\,s, because \kira{} and \pyred{} use different algorithms for the forward elimination and hence produce different triangular systems.
Both increases are still completely irrelevant compared to the time saved by using the forward-solved system.

Thus, one should check whether the forward elimination can be computed algebraically to speed up the calculation with \firefly{}.
However, usually the coefficients of the system are considerably more difficult after the forward elimination and are more expensive to parse and evaluate.
This can completely offset the gain, especially for multi-scale problems.

\subsection{Double-pentagon topology in five-light-parton scattering}
\label{sect:double_pentagon}

\begin{figure}[ht]
\centering
\includegraphics{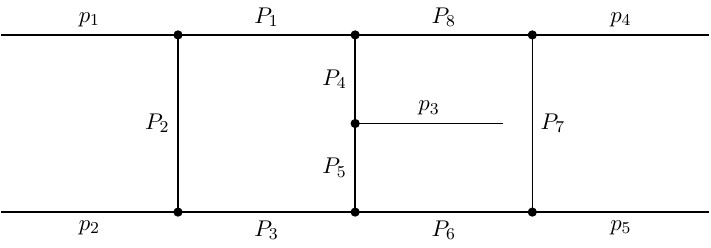}
\caption{The double-pentagon topology for five-light-parton scattering. The propagators $P_i$ are defined in Eq.\,\eqref{eq:dp_prop}.}
\label{fig:dp}
\end{figure}

The double-pentagon topology that appears in the amplitude of five-light-parton scattering at the two-loop level is illustrated in Fig.~\ref{fig:dp}.
There are five external momenta $p_1,\dots,p_5$ fulfilling $p_i^2=0$. All $p_i$ are assumed to be incoming, i.e.\ $\sum_i p_i=0$.
The kinematical invariants are defined by
\begin{equation}
  \begin{gathered}
    s_{12} = (p_1+p_2)^2,\quad s_{23} = (p_2+p_3)^2,\quad s_{34} = (p_3+p_4)^2,\\
    s_{45} = (p_4+p_5)^2,\quad s_{51} = (p_5+p_1)^2.
  \end{gathered}
\end{equation}
Including $d$, the reduction of the double-pentagon topology is a six variable problem. By setting $s_{12} = 1$ and restoring its dependence by dimensional analysis after the reduction, it can be reduced to a five variable problem.

The set of denominators describing the topology depicted in Fig.~\ref{fig:dp} is chosen as
\begin{equation}
  \begin{gathered}
    P_1 = l_1^2,\quad
    P_2 = (l_1 + p_1)^2,\quad
    P_3 = (l_1 + p_1 + p_2)^2,\quad
    P_4 = l_2^2,\\
    P_5 = (l_2 + p_3)^2,\quad
    P_6 = (l_1 + l_2 + p_1 + p_2 + p_3)^2,\quad
    P_7 = (l_1 + l_2 - p_4)^2,\\
    P_8 = (l_1 + l_2)^2,\quad
    P_9 = (l_2 + p_1)^2,\quad
    P_{10} = (l_2 + p_2)^2,\quad
    P_{11} = (l_2 + p_4)^2,
  \end{gathered}
  \label{eq:dp_prop}
\end{equation}
where the last three entries are auxiliary denominators. The system of equations used in this reduction is taken from Ref.~\cite{Guan:2019bcx}, which provides a system in block-triangular form.\footnote{Note that Ref.~\cite{Guan:2019bcx} uses a different propagator definition than given in their ancillary files. The definition in Eq.\,\eqref{eq:dp_prop} matches the one given in the ancillary files.}
This form is much better suited for the reduction than a naive IBP system as generated, e.g., by \kira{}.
The coefficients appearing in the system are further processed by \code{FireFly} to be cast in Horner form to optimize the evaluations.

We benchmark the reduction of all integrals including five scalar products. The reductions are performed on a machine with two Intel Xeon Gold 6138 with 20 cores each and 768\,GiB of RAM in total with hyperthreading enabled. All cores including hyper-threading are utilised with the option \code{-{}-parallel=80}. The integral selection is done using the option \code{select\_mandatory\_list} and we perform a numerical interpolation, i.e.\ \code{run\_firefly} is set to \code{true}. Hence, $2268$ master integral coefficients have to be interpolated. Additionally, we set the maximum bunch size to $128$. Further simplifications are obtained by \code{FireFly}'s factor scan. In total $19222$ factors can be found of which $1930$ are attributed to $s_{23}$, $2176$ to $s_{34}$, $2046$ to $s_{45}$, $2306$ to $s_{51}$, and $10764$ to $d$.
The results of this benchmark are shown in Tab.~\ref{tab:dp_bench}.
\begin{table}[ht]
  \begin{center}
    \caption{Benchmark results for the reduction of the double-pentagon topology with the configuration described in the main text.}
    \label{tab:dp_bench}
    {\renewcommand{\arraystretch}{1.3}
    \begin{tabular}{c c c c c}
      \toprule
      Runtime & Memory & Probes & \begin{tabular}{@{}c@{}}CPU time \\ per probe\end{tabular} & \begin{tabular}{@{}c@{}}CPU time \\ for probes\end{tabular}\\
      \midrule
      12\,d & 540\,GiB & 38278000 & 0.37\,s & 25\,\%\\
      \bottomrule
    \end{tabular}}
  \end{center}
\end{table}
The most complicated master integral coefficient has a maximum degree in the numerator of $87$ and in the denominator of $50$ thus yielding a dense bound of roughly $5.46 \cdot 10^7$ possible non-zero monomials. Fortunately, many of these are zero such that only about $10^7$ monomials contribute. Without the scan for factors, the maximum degree of the denominator of the most complicated coefficient rises to $85$. The database of the reduction occupies $25$\,GiB of disk space.

Although the number of required probes is comparably high, they can be computed relatively fast due to the block-triangular structure obtained in Ref.~\cite{Guan:2019bcx}. However, the interpolation of each coefficient is performed on a single thread and can become expensive in terms of runtime as the number of monomials usually rises factorially with the number of variables. Hence, in this example, the runtime for the interpolation is dominating. Note that there are two strategies to reduce the used memory of this calculation. On the one hand, the maximum bunch size can be reduced. On the other hand, the option \code{iterative\_reduction} with \code{masterwise} or \code{sectorwise} can be used. By employing the latter option, the calculation can be distributed manually (sector- or masterwise) on several machines to obtain a runtime improvement in addition.

\section{Installation}
\label{sect:installation}

\subsection{Obtaining Kira}

A statically linked executable of \kira{} for \code{Linux x86\_64} is available from our web page at \url{https://kira.hepforge.org}.
This executable has all optional features included except for \code{MPI}.
If you require \code{MPI}, you must compile \kira{} yourself against the \code{MPI} version used on your computer cluster.

The source code of \kira{} is available from our \git{} repository at
\gitlab{} under the URL \url{https://gitlab.com/kira-pyred/kira}.
To obtain the source code of the latest release version, clone the repository
with
\begin{verbatim}
  git clone https://gitlab.com/kira-pyred/kira.git -b release
\end{verbatim}
checking out the release branch.
Release versions are also available as \git{} tags (e.g.\ \code{kira-2.0}).
The master branch of the repository contains the latest pre-release version,
receiving more frequent updates with new features and fixes.
To obtain the source code of the latest pre-release version, clone the
repository with
\begin{verbatim}
  git clone https://gitlab.com/kira-pyred/kira.git
\end{verbatim}
checking out the master branch.
Packages of the release versions as compressed \code{Tar} archives are
available from \url{https://kira.hepforge.org}.

\subsection{Prerequisites}
\label{sect:prerequisites}

\noindent
\textit{Platform requirements}\smallskip\\
\code{Linux x86\_64} or \code{macOS}.

\medskip\pagebreak[2]
\noindent
\textit{Compiler requirements}\smallskip\\
A \code{C++} compiler supporting the \code{C++14} standard and a \code{C}
compiler supporting the \code{C11} standard.

\medskip\pagebreak[2]
\noindent
\textit{Build system requirements}\smallskip\\
\kira{} can either be built with the \code{Meson} build system~\cite{Meson}
version \code{0.46} or later and \code{Ninja}~\cite{Ninja},
or with the \code{Autotools} build system~\cite{Autotools}.

We recommend to use the \code{Meson} build system.
If \code{Meson} is not available on your system, it can be installed into your home directory as non-root user with
\begin{verbatim}
  pip3 install --user meson
\end{verbatim}
(requires \code{Python}\;\code{3.5} or later).
This will install \code{Meson} to \code{\textasciitilde/.local/bin}.

The \code{Ninja} binary (and source code) is available from
\url{https://ninja-build.org}.

\medskip\pagebreak[2]
\noindent
\textit{Dependencies}\smallskip\\
\kira{} requires the following packages to be installed on the system:
\begin{itemize}
  \item \code{GiNaC}~\cite{GiNaC,Bauer:2000cp,Vollinga:2005pk}, which itself
    requires \code{CLN}~\cite{CLN},
  \item \code{zlib}~\cite{ZLIB}.
  \item \code{Fermat}~\cite{Fermat} is required to run \kira{}.
\end{itemize}
If the \code{Fermat} executable is not found automatically at startup,
or a specific \code{Fermat} installation should be used, the path to the
\code{Fermat} executable can be provided via the environment variable
\code{FERMATPATH}.

Depending on the enabled optional features of \kira{}, the following packages
are required in addition:
\begin{itemize}
  \item \code{GMP} \cite{GMP} if \firefly{} is used,
  \item \code{MPFR} \cite{MPFR} if \flint{} is used,
  \item an \code{MPI} \cite{mpi_forum} library (disabled by default) for
    parallelization on computer clusters,
  \item \code{jemalloc} \cite{jemalloc} (disabled by default) for more efficient
    memory allocation.
\end{itemize}
The following dependencies can be automatically built and installed as
subprojects with the \code{Meson} build system, i.e.\ if they are not found on
the system, they will be built automatically along with \kira{}: %
\begin{itemize}
  \item \code{yaml-cpp}~\cite{YAMLCPP} (required),
  \item \firefly{} \cite{Klappert:2020aqs} (optional, enabled by default).
    If you decide to install \firefly{} manually, we recommend to use the
    version from the branch \code{kira-2} of the \code{Git} repository at
    \url{https://gitlab.com/firefly-library/firefly}.
    This branch will remain compatible with \kiraZO{}.
  \item \flint{} \cite{flint} (optional, enabled by default).
    We recommend using \flint{}, because it not only offers better performance
    for the finite field arithmetic, but is also required to enable some
    features of \code{FireFly}, most notably the factor scan.
\end{itemize}
If the \code{Autotools} build system is used, all enabled dependencies must be
installed manually.
If \code{FireFly} is not build as a subproject, to use \flint{} and \code{MPI},
they must be enabled in \code{FireFly}'s \code{CMake} build system.

Note that \code{GiNaC}, \code{CLN}, \code{yaml-cpp}, and \firefly{}
must have been compiled with the same compiler which is used to compile \kira{}.
Otherwise the linking step will most likely fail.
If you are using the system compiler, you can usually install \code{GiNaC},
\code{CLN}, and \code{yaml-cpp} via your system's package manager.
However, if you are using a different compiler, this usually means in practice
that you also have to build these packages from source and, if installed
with a non-default installation prefix, the environment variables
\code{C\_PATH}, \code{LD\_LIBRARY\_PATH} and \code{PKG\_CONFIG\_PATH}
must be set accordingly.

\subsection{Compiling Kira with the Meson build system}

To build \kira{} with the \code{Meson} build system, \code{Meson}\;\code{0.46}
(or later) and \code{Ninja} are required.
If \code{Meson} and \code{Ninja} are not available on your system, see
paragraph \textit{``Build system requirements''} in
Sect.~\ref{sect:prerequisites}.

To compile and install \kira{}, run
\begin{verbatim}
  meson --prefix=/install/path builddir
  cd builddir
  ninja
  ninja install
\end{verbatim}
where \code{builddir} is the build directory.
Specifying the installation prefix with \code{-{}-prefix} is optional.

\medskip\pagebreak[2]
\noindent
\textit{Build options}
%
\begin{itemize}
  \item \code{-Dfirefly=false} (default: \code{true}):
    Build without \code{FireFly} support.
  \item \code{-Dflint=false} (default: \code{true}):
    If \code{FireFly} is built as a subproject, disable \flint{}.
  \item \code{-Dmpi=true} (default: \code{false}):
    If \code{FireFly} is built as a subproject, enable \code{MPI}.
    This is known to work best with \code{OpenMPI}~\cite{openmpi,Gabriel:2004}.
    For performance reasons, we recommend \code{MPICH}~\cite{mpich}, though.
  \item \code{-Dcustom-mpi=<name>}: If your \code{MPI} installation provides
    a \code{pkg-config} file, but is not found automatically with
    \code{-Dmpi=true}, pass the name of the \code{MPI} implementation
    as \code{<name>}, e.g.\ \code{-Dcustom-mpi=mpich}.
    Some systems don't provide a \code{pkg-config} file for \code{MPICH}.
    In that case we recommend to install \code{FireFly} with its own
    \code{CMake} build system instead.
  \item \code{-Djemalloc=true} (default: \code{false}):
    Link with the \code{jemalloc} memory allocator~\cite{jemalloc}.
    This can lead to significantly increased performance, often by more
    than $20\,\%$ from our experience if \firefly{} is used.
    However, using \code{jemalloc} may not work on some systems, especially in
    combination with certain \code{MPI} implementations.\footnote{
    This can depend on subtleties like the linking order of the \code{jemalloc}
    and \code{MPI} libraries.}
    Alternatively, to use \code{jemalloc}, one can set the
    environment variable \code{LD\_PRELOAD} to point to \code{jemalloc.so}
    and export it.
\end{itemize}
To show the full list of available build options, run
\code{meson}\;\code{configure} in the build directory.

\medskip\pagebreak[2]
\noindent
\textit{Subprojects}\smallskip\\
%
If \code{yaml-cpp} or \code{FireFly} are not found on the system, per default
they will be downloaded and built as \code{Meson} subprojects.
If the option \code{-Dflint=true} (default) is set and \code{FireFly} is built
as a subproject, also \flint{} will be downloaded and built as a subproject
if it is not found on the system.

The usage of subprojects can be controlled with the following options:
\begin{itemize}
  \item \code{-{}-wrap-mode=nodownload}:
    Do not download subprojects, but build them if already available
    (and not found on the system).
  \item \code{-{}-wrap-mode=nofallback}:
    Do not build subprojects, even if the libraries are not found on the system.
  \item \code{-{}-wrap-mode=forcefallback}:
    Build subprojects even if the libraries can be found on the system.
  \item \code{-{}-force-fallback-for=<deps>}:
    Like \code{forcefallback}, but only for dependencies in the comma separated
    list \code{<deps>}. Overrides \code{nofallback} and \code{forcefallback}.
\end{itemize}
These options are only fully supported with \code{Meson}\;\code{0.49} or later.
For details see\\
\url{https://mesonbuild.com/Subprojects.html}.

\textit{Note:}
Subprojects are not updated automatically.
To update subprojects, run
\begin{verbatim}
  meson subprojects update
\end{verbatim}
(requires \code{Meson}\;\code{0.49} or later).
\code{Git} subprojects can of course also be manually updated by running
\code{git}\;\code{pull} in the corresponding subproject directory
(e.g.\ \code{subprojects/firefly}).

\subsection{Compiling Kira with the Autotools build system}
\label{installwithautotools}

First run
\begin{verbatim}
  autoreconf -i
\end{verbatim}
and then compile and install with
\begin{verbatim}
  ./configure --prefix=/install/path --enable-firefly=yes
  make
  make install
\end{verbatim}
where the optional \code{-{}-prefix} argument sets the installation prefix.
Without the option \code{-{}-enable-firefly=yes}, \kira{} will be built without
\firefly{} support.
Note that subproject installation is not supported with the \code{Autotools}
build system, i.e.\ all dependencies must be installed manually.

\section{Conclusions}
\label{sect:conclusions}

In this article we presented the new version \code{2.0} of the Feynman integral reduction program \kira{}.
The major new features introduced in this release are the reconstruction of final coefficients by means of finite field methods with the help of \firefly{}, and the parallelization of this procedure on computer clusters with \mpi{}.
Besides many minor improvements and extensions, the framework for the solution of user-provided systems of equations has been extended to support most features available for integration-by-parts reductions.

We reproduced benchmarks from previous \kira{} publications, showing significantly increased performance and reduced main memory consumption.
Furthermore we provide new state-of-the-art benchmark points to demonstrate the effect of various newly added features on the computing resource requirements.

\section*{Acknowledgments}

The research of J.K.\ and F.L.\ was supported by the \textit{Deutsche Forschungsgemeinschaft} (DFG, German Research Foundation) under grant \href{http://gepris.dfg.de/gepris/projekt/396021762?language=en}{396021762} -- \href{http://p3h.particle.kit.edu/start}{TRR 257}.
Furthermore, F.L.\ acknowledges financial support by the DFG through project \href{http://gepris.dfg.de/gepris/projekt/386986591?language=en}{386986591}.
J.U.\ received funding in the early stage of this work from the European Research Council (ERC) under the European Union’s Horizon 2020 research and innovation programme under grant agreement no.\ 647356 (CutLoops).
This work has been supported by the Mainz Institute for Theoretical Physics (MITP) of the Cluster of Excellence PRISMA+ (Project ID 39083149).
Some of the calculations were performed with computing resources granted by RWTH Aachen University under project rwth0541.

The Feynman diagrams in this paper were drawn with Ti\textit{k}Z-Feynman\,\cite{Ellis:2016jkw}.

\appendix


\bibliographystyle{JHEP}
\bibliography{literature}{}

\providecommand{\href}[2]{#2}\begingroup\raggedright\begin{thebibliography}{10}

\bibitem{Amoroso:2020lgh}
S.~Amoroso et~al., \emph{{Les Houches 2019: Physics at TeV Colliders: Standard
  Model Working Group Report}},  in \emph{{11th Les Houches Workshop on Physics
  at TeV Colliders}: {PhysTeV Les Houches}}, 3, 2020,
  \href{https://arxiv.org/abs/2003.01700}{{\ttfamily 2003.01700}}.

\bibitem{Henn:2013pwa}
J.~M. Henn, \emph{{Multiloop Integrals in Dimensional Regularization Made
  Simple}}, \href{https://doi.org/10.1103/PhysRevLett.110.251601}{\emph{Phys.
  Rev. Lett.} {\bfseries 110} (2013) 251601}
  [\href{https://arxiv.org/abs/1304.1806}{{\ttfamily 1304.1806}}].

\bibitem{Argeri:2014qva}
M.~Argeri, S.~Di~Vita, P.~Mastrolia, E.~Mirabella, J.~Schlenk, U.~Schubert
  et~al., \emph{{Magnus and Dyson series for Master Integrals}},
  \href{https://doi.org/10.1007/JHEP03(2014)082}{\emph{JHEP} {\bfseries 03}
  (2014) 082} [\href{https://arxiv.org/abs/1401.2979}{{\ttfamily 1401.2979}}].

\bibitem{vonManteuffel:2015gxa}
A.~von Manteuffel, E.~Panzer and R.~M. Schabinger, \emph{{Computation of form
  factors in massless QCD with finite master integrals}},
  \href{https://doi.org/10.1103/PhysRevD.93.125014}{\emph{Phys. Rev. D}
  {\bfseries 93} (2016) 125014}
  [\href{https://arxiv.org/abs/1510.06758}{{\ttfamily 1510.06758}}].

\bibitem{Francesco:2019yqt}
F.~Moriello, \emph{{Generalised power series expansions for the elliptic planar
  families of Higgs + jet production at two loops}},
  \href{https://doi.org/10.1007/JHEP01(2020)150}{\emph{JHEP} {\bfseries 01}
  (2020) 150} [\href{https://arxiv.org/abs/1907.13234}{{\ttfamily
  1907.13234}}].

\bibitem{Hidding:2020ytt}
M.~Hidding, \emph{{DiffExp, a Mathematica package for computing Feynman
  integrals in terms of one-dimensional series expansions}},
  \href{https://arxiv.org/abs/2006.05510}{{\ttfamily 2006.05510}}.

\bibitem{Gluza:2010ws}
J.~Gluza, K.~Kajda and D.~A. Kosower, \emph{{Towards a basis for planar
  two-loop integrals}},
  \href{https://doi.org/10.1103/PhysRevD.83.045012}{\emph{Phys. Rev.}
  {\bfseries D83} (2011) 045012}
  [\href{https://arxiv.org/abs/1009.0472}{{\ttfamily 1009.0472}}].

\bibitem{Schabinger:2011dz}
R.~M. Schabinger, \emph{{A new algorithm for the generation of
  unitarity-compatible integration by parts relations}},
  \href{https://doi.org/10.1007/JHEP01(2012)077}{\emph{JHEP} {\bfseries 01}
  (2012) 077} [\href{https://arxiv.org/abs/1111.4220}{{\ttfamily 1111.4220}}].

\bibitem{Ita:2015tya}
H.~Ita, \emph{{Two-loop integrand decomposition into master integrals and
  surface terms}},
  \href{https://doi.org/10.1103/PhysRevD.94.116015}{\emph{Phys. Rev.}
  {\bfseries D94} (2016) 116015}
  [\href{https://arxiv.org/abs/1510.05626}{{\ttfamily 1510.05626}}].

\bibitem{Boehm:2017wjc}
J.~Böhm, A.~Georgoudis, K.~J. Larsen, M.~Schulze and Y.~Zhang, \emph{{Complete
  sets of logarithmic vector fields for integration-by-parts identities of
  Feynman integrals}},
  \href{https://doi.org/10.1103/PhysRevD.98.025023}{\emph{Phys. Rev.}
  {\bfseries D98} (2018) 025023}
  [\href{https://arxiv.org/abs/1712.09737}{{\ttfamily 1712.09737}}].

\bibitem{Kosower:2018obg}
D.~A. Kosower, \emph{{Direct solution of integration-by-parts systems}},
  \href{https://doi.org/10.1103/PhysRevD.98.025008}{\emph{Phys. Rev.}
  {\bfseries D98} (2018) 025008}
  [\href{https://arxiv.org/abs/1804.00131}{{\ttfamily 1804.00131}}].

\bibitem{vonManteuffel:2020vjv}
A.~von Manteuffel, E.~Panzer and R.~M. Schabinger, \emph{{Cusp and Collinear
  Anomalous Dimensions in Four-Loop QCD from Form Factors}},
  \href{https://doi.org/10.1103/PhysRevLett.124.162001}{\emph{Phys. Rev. Lett.}
  {\bfseries 124} (2020) 162001}
  [\href{https://arxiv.org/abs/2002.04617}{{\ttfamily 2002.04617}}].

\bibitem{Larsen:2015ped}
K.~J. Larsen and Y.~Zhang, \emph{{Integration-by-parts reductions from
  unitarity cuts and algebraic geometry}},
  \href{https://doi.org/10.1103/PhysRevD.93.041701}{\emph{Phys. Rev.}
  {\bfseries D93} (2016) 041701}
  [\href{https://arxiv.org/abs/1511.01071}{{\ttfamily 1511.01071}}].

\bibitem{Boehm:2018fpv}
J.~Böhm, A.~Georgoudis, K.~J. Larsen, H.~Schönemann and Y.~Zhang,
  \emph{{Complete integration-by-parts reductions of the non-planar hexagon-box
  via module intersections}},
  \href{https://doi.org/10.1007/JHEP09(2018)024}{\emph{JHEP} {\bfseries 09}
  (2018) 024} [\href{https://arxiv.org/abs/1805.01873}{{\ttfamily
  1805.01873}}].

\bibitem{Bendle:2019csk}
D.~Bendle, J.~Böhm, W.~Decker, A.~Georgoudis, F.-J. Pfreundt, M.~Rahn et~al.,
  \emph{{Integration-by-parts reductions of Feynman integrals using Singular
  and GPI-Space}}, \href{https://doi.org/10.1007/JHEP02(2020)079}{\emph{JHEP}
  {\bfseries 02} (2020) 079}
  [\href{https://arxiv.org/abs/1908.04301}{{\ttfamily 1908.04301}}].

\bibitem{Mastrolia:2018uzb}
P.~Mastrolia and S.~Mizera, \emph{{Feynman integrals and intersection theory}},
  \href{https://doi.org/10.1007/JHEP02(2019)139}{\emph{JHEP} {\bfseries 02}
  (2019) 139} [\href{https://arxiv.org/abs/1810.03818}{{\ttfamily
  1810.03818}}].

\bibitem{Frellesvig:2019kgj}
H.~Frellesvig, F.~Gasparotto, S.~Laporta, M.~K. Mandal, P.~Mastrolia,
  L.~Mattiazzi et~al., \emph{{Decomposition of Feynman integrals on the maximal
  cut by intersection numbers}},
  \href{https://doi.org/10.1007/JHEP05(2019)153}{\emph{JHEP} {\bfseries 05}
  (2019) 153} [\href{https://arxiv.org/abs/1901.11510}{{\ttfamily
  1901.11510}}].

\bibitem{Frellesvig:2019uqt}
H.~Frellesvig, F.~Gasparotto, M.~K. Mandal, P.~Mastrolia, L.~Mattiazzi and
  S.~Mizera, \emph{{Vector Space of Feynman Integrals and Multivariate
  Intersection Numbers}},
  \href{https://doi.org/10.1103/PhysRevLett.123.201602}{\emph{Phys. Rev. Lett.}
  {\bfseries 123} (2019) 201602}
  [\href{https://arxiv.org/abs/1907.02000}{{\ttfamily 1907.02000}}].

\bibitem{Weinzierl:2020xyy}
S.~Weinzierl, \emph{{On the computation of intersection numbers for twisted
  cocycles}},  \href{https://arxiv.org/abs/2002.01930}{{\ttfamily 2002.01930}}.

\bibitem{Frellesvig:2020qot}
H.~Frellesvig, F.~Gasparotto, S.~Laporta, M.~K. Mandal, P.~Mastrolia,
  L.~Mattiazzi et~al., \emph{{Decomposition of Feynman Integrals by
  Multivariate Intersection Numbers}},
  \href{https://arxiv.org/abs/2008.04823}{{\ttfamily 2008.04823}}.

\bibitem{vonManteuffel:2014ixa}
A.~von Manteuffel and R.~M. Schabinger, \emph{{A novel approach to integration
  by parts reduction}},
  \href{https://doi.org/10.1016/j.physletb.2015.03.029}{\emph{Phys. Lett.}
  {\bfseries B744} (2015) 101}
  [\href{https://arxiv.org/abs/1406.4513}{{\ttfamily 1406.4513}}].

\bibitem{Peraro:2016wsq}
T.~Peraro, \emph{{Scattering amplitudes over finite fields and multivariate
  functional reconstruction}},
  \href{https://doi.org/10.1007/JHEP12(2016)030}{\emph{JHEP} {\bfseries 12}
  (2016) 030} [\href{https://arxiv.org/abs/1608.01902}{{\ttfamily
  1608.01902}}].

\bibitem{Smirnov:2019qkx}
A.~V. Smirnov and F.~S. Chukharev, \emph{{FIRE6: Feynman Integral REduction
  with modular arithmetic}},
  \href{https://doi.org/10.1016/j.cpc.2019.106877}{\emph{Comput. Phys. Commun.}
  {\bfseries 247} (2020) 106877}
  [\href{https://arxiv.org/abs/1901.07808}{{\ttfamily 1901.07808}}].

\bibitem{Klappert:2019emp}
J.~Klappert and F.~Lange, \emph{{Reconstructing rational functions with
  FireFly}}, \href{https://doi.org/10.1016/j.cpc.2019.106951}{\emph{Comput.
  Phys. Commun.} {\bfseries 247} (2020) 106951}
  [\href{https://arxiv.org/abs/1904.00009}{{\ttfamily 1904.00009}}].

\bibitem{Peraro:2019svx}
T.~Peraro, \emph{{FiniteFlow: multivariate functional reconstruction using
  finite fields and dataflow graphs}},
  \href{https://doi.org/10.1007/JHEP07(2019)031}{\emph{JHEP} {\bfseries 07}
  (2019) 031} [\href{https://arxiv.org/abs/1905.08019}{{\ttfamily
  1905.08019}}].

\bibitem{Klappert:2020aqs}
J.~Klappert, S.~Y. Klein and F.~Lange, \emph{{Interpolation of dense and sparse
  rational functions and other improvements in FireFly}},
  \href{https://doi.org/10.1016/j.cpc.2021.107968}{\emph{Comput. Phys. Commun.}
  {\bfseries 264} (2021) 107968}
  [\href{https://arxiv.org/abs/2004.01463}{{\ttfamily 2004.01463}}].

\bibitem{Liu:2017jxz}
X.~Liu, Y.-Q. Ma and C.-Y. Wang, \emph{{A systematic and efficient method to
  compute multi-loop master integrals}},
  \href{https://doi.org/10.1016/j.physletb.2018.02.026}{\emph{Phys. Lett.}
  {\bfseries B779} (2018) 353}
  [\href{https://arxiv.org/abs/1711.09572}{{\ttfamily 1711.09572}}].

\bibitem{Liu:2018dmc}
X.~Liu and Y.-Q. Ma, \emph{{Determining arbitrary Feynman integrals by vacuum
  integrals}}, \href{https://doi.org/10.1103/PhysRevD.99.071501}{\emph{Phys.
  Rev.} {\bfseries D99} (2019) 071501}
  [\href{https://arxiv.org/abs/1801.10523}{{\ttfamily 1801.10523}}].

\bibitem{Wang:2019mnn}
Y.~Wang, Z.~Li and N.~ul~Basat, \emph{{Direct reduction of multiloop multiscale
  scattering amplitudes}},
  \href{https://doi.org/10.1103/PhysRevD.101.076023}{\emph{Phys. Rev. D}
  {\bfseries 101} (2020) 076023}
  [\href{https://arxiv.org/abs/1901.09390}{{\ttfamily 1901.09390}}].

\bibitem{Guan:2019bcx}
X.~Guan, X.~Liu and Y.-Q. Ma, \emph{{Complete reduction of integrals in
  two-loop five-light-parton scattering amplitudes}},
  \href{https://doi.org/10.1088/1674-1137/44/9/093106}{\emph{Chin. Phys. C}
  {\bfseries 44} (2020) 093106}
  [\href{https://arxiv.org/abs/1912.09294}{{\ttfamily 1912.09294}}].

\bibitem{Laporta:2001dd}
S.~Laporta, \emph{{High-precision calculation of multiloop Feynman integrals by
  difference equations}},
  \href{https://doi.org/10.1016/S0217-751X(00)00215-7}{\emph{Int.J.Mod.Phys.}
  {\bfseries A15} (2000) 5087}
  [\href{https://arxiv.org/abs/hep-ph/0102033}{{\ttfamily hep-ph/0102033}}].

\bibitem{Anastasiou:2004vj}
C.~Anastasiou and A.~Lazopoulos, \emph{{Automatic integral reduction for higher
  order perturbative calculations}},
  \href{https://doi.org/10.1088/1126-6708/2004/07/046}{\emph{JHEP} {\bfseries
  07} (2004) 046} [\href{https://arxiv.org/abs/hep-ph/0404258}{{\ttfamily
  hep-ph/0404258}}].

\bibitem{vonManteuffel:2012np}
A.~von Manteuffel and C.~Studerus, \emph{{Reduze 2 -- Distributed Feynman
  Integral Reduction}},  \href{https://arxiv.org/abs/1201.4330}{{\ttfamily
  1201.4330}}.

\bibitem{Smirnov:2014hma}
A.~V. Smirnov, \emph{{FIRE5: A C++ implementation of Feynman Integral
  REduction}}, \href{https://doi.org/10.1016/j.cpc.2014.11.024}{\emph{Comput.
  Phys. Commun.} {\bfseries 189} (2015) 182}
  [\href{https://arxiv.org/abs/1408.2372}{{\ttfamily 1408.2372}}].

\bibitem{Maierhoefer:2017hyi}
P.~Maierhöfer, J.~Usovitsch and P.~Uwer, \emph{{Kira—A Feynman integral
  reduction program}},
  \href{https://doi.org/10.1016/j.cpc.2018.04.012}{\emph{Comput. Phys. Commun.}
  {\bfseries 230} (2018) 99}
  [\href{https://arxiv.org/abs/1705.05610}{{\ttfamily 1705.05610}}].

\bibitem{mpi_forum}
{MPI Forum}, \emph{{Message Passing Interface}},
  \href{https://www.mpi-forum.org}{https://www.mpi-forum.org}.

\bibitem{Harlander:2018zpi}
R.~V. Harlander, Y.~Kluth and F.~Lange, \emph{{The two-loop energy–momentum
  tensor within the gradient-flow formalism}},
  \href{https://doi.org/10.1140/epjc/s10052-018-6415-7}{\emph{Eur. Phys. J.}
  {\bfseries C78} (2018) 944}
  [\href{https://arxiv.org/abs/1808.09837}{{\ttfamily 1808.09837}}].

\bibitem{Artz:2019bpr}
J.~Artz, R.~V. Harlander, F.~Lange, T.~Neumann and M.~Prausa, \emph{{Results
  and techniques for higher order calculations within the gradient-flow
  formalism}}, \href{https://doi.org/10.1007/JHEP06(2019)121}{\emph{JHEP}
  {\bfseries 06} (2019) 121}
  [\href{https://arxiv.org/abs/1905.00882}{{\ttfamily 1905.00882}}].

\bibitem{Harlander:2020duo}
R.~V. Harlander, F.~Lange and T.~Neumann, \emph{{Hadronic vacuum polarization
  using gradient flow}},
  \href{https://doi.org/10.1007/JHEP08(2020)109}{\emph{JHEP} {\bfseries 08}
  (2020) 109} [\href{https://arxiv.org/abs/2007.01057}{{\ttfamily
  2007.01057}}].

\bibitem{Tkachov:1981wb}
F.~V. Tkachov, \emph{{A theorem on analytical calculability of 4-loop
  renormalization group functions}},
  \href{https://doi.org/10.1016/0370-2693(81)90288-4}{\emph{Phys. Lett. B}
  {\bfseries 100} (1981) 65}.

\bibitem{Chetyrkin:1981qh}
K.~G. Chetyrkin and F.~V. Tkachov, \emph{{Integration by parts: The algorithm
  to calculate $\beta$-functions in 4 loops}},
  \href{https://doi.org/10.1016/0550-3213(81)90199-1}{\emph{Nucl. Phys.}
  {\bfseries B192} (1981) 159}.

\bibitem{Gehrmann:1999as}
T.~Gehrmann and E.~Remiddi, \emph{{Differential equations for two-loop
  four-point functions}},
  \href{https://doi.org/10.1016/S0550-3213(00)00223-6}{\emph{Nucl. Phys.}
  {\bfseries B580} (2000) 485}
  [\href{https://arxiv.org/abs/hep-ph/9912329}{{\ttfamily hep-ph/9912329}}].

\bibitem{Fermat}
R.~H. Lewis, \emph{{Computer Algebra System Fermat}},
  \href{https://home.bway.net/lewis}{https://home.bway.net/lewis}.

\bibitem{Kauers:2008zz}
M.~Kauers, \emph{{Fast Solvers for Dense Linear Systems}},
  \href{https://doi.org/10.1016/j.nuclphysbps.2008.09.111}{\emph{Nucl. Phys. B
  Proc. Suppl.} {\bfseries 183} (2008) 245}.

\bibitem{Kant:2013vta}
P.~Kant, \emph{{Finding linear dependencies in integration-by-parts equations:
  A Monte Carlo approach}},
  \href{https://doi.org/10.1016/j.cpc.2014.01.017}{\emph{Comput. Phys. Commun.}
  {\bfseries 185} (2014) 1473}
  [\href{https://arxiv.org/abs/1309.7287}{{\ttfamily 1309.7287}}].

\bibitem{deKleine:2005}
J.~de~Kleine, M.~Monagan and A.~Wittkopf, \emph{{Algorithms for the Non-monic
  Case of the Sparse Modular GCD Algorithm}},
  \href{https://doi.org/10.1145/1073884.1073903}{\emph{Proc. Int. Symp.
  Symbolic Algebraic Comp.} {\bfseries 2005} (2005) 124}.

\bibitem{Zippel:1979}
R.~Zippel, \emph{{Probabilistic algorithms for sparse polynomials}},
  \href{https://doi.org/10.1007/3-540-09519-5_73}{\emph{Symbolic Algebraic
  Comp. EUROSAM} {\bfseries 1979} (1979) 216}.

\bibitem{Ben-Or:1988}
M.~Ben-Or and P.~Tiwari, \emph{{A Deterministic Algorithm for Sparse
  Multivariate Polynomial Interpolation}},
  \href{https://doi.org/10.1145/62212.62241}{\emph{Proc. ACM Symp. Theory
  Comp.} {\bfseries 20} (1988) 301}.

\bibitem{Kaltofen:1988}
{E. Kaltofen and Lakshman Y.}, \emph{{Improved Sparse Multivariate Polynomial
  Interpolation Algorithms}},
  \href{https://doi.org/10.1007/3-540-51084-2_44}{\emph{Symbolic Algebraic
  Comp. ISSAC} {\bfseries 1988} (1989) 467}.

\bibitem{Zippel:1990}
R.~Zippel, \emph{{Interpolating Polynomials from their Values}},
  \href{https://doi.org/10.1016/S0747-7171(08)80018-1}{\emph{J. Symb. Comp.}
  {\bfseries 9} (1990) 375}.

\bibitem{Kaltofen:2000}
{E. Kaltofen, W.-s. Lee and A. A. Lobo}, \emph{{Early Termination in
  Ben-Or/Tiwari Sparse Interpolation and a Hybrid of Zippel's Algorithm}},
  \href{https://doi.org/10.1145/345542.345629}{\emph{Proc. Int. Symp. Symbolic
  Algebraic Comp.} {\bfseries 2000} (2000) 192}.

\bibitem{Kaltofen:2003}
{E. Kaltofen and W.-s. Lee}, \emph{{Early termination in sparse interpolation
  algorithms}}, \href{https://doi.org/10.1016/S0747-7171(03)00088-9}{\emph{J.
  Symb. Comp.} {\bfseries 36} (2003) 365}.

\bibitem{Cuyt:2011}
{A. Cuyt and W.-s. Lee}, \emph{{Sparse interpolation of multivariate rational
  functions}}, \href{https://doi.org/10.1016/j.tcs.2010.11.050}{\emph{Theor.
  Comp. Sci.} {\bfseries 412} (2011) 1445}.

\bibitem{Wang:1981}
P.~S. Wang, \emph{{A p-adic Algorithm for Univariate Partial Fractions}},
  \href{https://doi.org/10.1145/800206.806398}{\emph{Proc. ACM Symp. Symbolic
  Algebraic Comp.} {\bfseries 1981} (1981) 212}.

\bibitem{Monagan:2004}
M.~Monagan, \emph{{Maximal Quotient Rational Reconstruction: An Almost Optimal
  Algorithm for Rational Reconstruction}},
  \href{https://doi.org/10.1145/1005285.1005321}{\emph{Proc. Int. Symp.
  Symbolic Algebraic Comp.} {\bfseries 2004} (2004) 243}.

\bibitem{von_zur_Gathen}
J.~von~zur Gathen and J.~Gerhard, \emph{{Modern Computer Algebra}}. {Cambridge
  University Press}, {third}~ed., 2013,
  \href{https://doi.org/10.1017/CBO9781139856065}{10.1017/CBO9781139856065}.

\bibitem{DeMillo:1978}
R.~A. DeMillo and R.~J. Lipton, \emph{{A probabilistic remark on algebraic
  program testing}},
  \href{https://doi.org/10.1016/0020-0190(78)90067-4}{\emph{Inf. Process.
  Lett.} {\bfseries 7} (1978) 193}.

\bibitem{Schwartz:1980}
J.~T. Schwartz, \emph{{Fast Probabilistic Algorithms for Verification of
  Polynomial Identities}},
  \href{https://doi.org/10.1145/322217.322225}{\emph{J. ACM} {\bfseries 27}
  (1980) 701}.

\bibitem{jemalloc}
J.~Evans et~al., \emph{{jemalloc memory allocator}},
  \href{http://jemalloc.net}{http://jemalloc.net}.

\bibitem{kira:1_1}
P.~Maierhöfer and J.~Usovitsch, \emph{{Kira 1.1 Release Notes}},
  \href{https://kira.hepforge.org/downloads?f=papers/kira-release-notes-1.1.pdf}{https://kira.hepforge.org/downloads?f=papers/kira-release-notes-1.1.pdf}.

\bibitem{Maierhofer:2018gpa}
P.~Maierhöfer and J.~Usovitsch, \emph{{Kira 1.2 Release Notes}},
  \href{https://arxiv.org/abs/1812.01491}{{\ttfamily 1812.01491}}.

\bibitem{BSMF_1992__120_3_371_0}
C.~Sabbah, \emph{Lieu des p\^oles d'un syst\`eme holonome d'\'equations aux
  diff\'erences finies},
  \href{https://doi.org/10.24033/bsmf.2191}{\emph{Bulletin de la Soci\'et\'e
  Math\'ematique de France} {\bfseries 120} (1992) 371}.

\bibitem{Smirnov:2020quc}
A.~V. Smirnov and V.~A. Smirnov, \emph{{How to choose master integrals}},
  \href{https://doi.org/10.1016/j.nuclphysb.2020.115213}{\emph{Nucl. Phys. B}
  {\bfseries 960} (2020) 115213}
  [\href{https://arxiv.org/abs/2002.08042}{{\ttfamily 2002.08042}}].

\bibitem{Usovitsch:2020jrk}
J.~Usovitsch, \emph{{Factorization of denominators in integration-by-parts
  reductions}},  \href{https://arxiv.org/abs/2002.08173}{{\ttfamily
  2002.08173}}.

\bibitem{Papadopoulos:2014hla}
C.~G. Papadopoulos, D.~Tommasini and C.~Wever, \emph{{Two-loop master integrals
  with the simplified differential equations approach}},
  \href{https://doi.org/10.1007/JHEP01(2015)072}{\emph{JHEP} {\bfseries 01}
  (2015) 072} [\href{https://arxiv.org/abs/1409.6114}{{\ttfamily 1409.6114}}].

\bibitem{Hidding:2017jkk}
M.~Hidding and F.~Moriello, \emph{{All orders structure and efficient
  computation of linearly reducible elliptic Feynman integrals}},
  \href{https://doi.org/10.1007/JHEP01(2019)169}{\emph{JHEP} {\bfseries 01}
  (2019) 169} [\href{https://arxiv.org/abs/1712.04441}{{\ttfamily
  1712.04441}}].

\bibitem{Papadopoulos:2019iam}
C.~G. Papadopoulos and C.~Wever, \emph{{Internal reduction method for computing
  Feynman integrals}},
  \href{https://doi.org/10.1007/JHEP02(2020)112}{\emph{JHEP} {\bfseries 02}
  (2020) 112} [\href{https://arxiv.org/abs/1910.06275}{{\ttfamily
  1910.06275}}].

\bibitem{Grozin:2004yc}
A.~G. Grozin, \emph{{Heavy Quark Effective Theory}},
  \href{https://doi.org/10.1007/b79301}{\emph{Springer Tracts Mod. Phys.}
  {\bfseries 201} (2004) 1}.

\bibitem{Frellesvig:2018ymi}
H.~Frellesvig, R.~Bonciani, V.~Del~Duca, F.~Moriello, J.~Henn and V.~Smirnov,
  \emph{{Non-planar two-loop Feynman integrals contributing to Higgs plus jet
  production}}, \href{https://doi.org/10.22323/1.303.0076}{\emph{PoS}
  {\bfseries LL2018} (2018) 076}.

\bibitem{intelmpi}
{Intel Corporation}, \emph{{Intel\,\textsuperscript{\textregistered} MPI
  Library}},
  \href{https://software.intel.com/content/www/us/en/develop/tools/mpi-library.html}{https://software.intel.com/content/www/us/en/develop/tools/mpi-library.html}.

\bibitem{Meson}
J.~Pakkanen, \emph{{The Meson Build system}},
  \href{https://mesonbuild.com}{https://mesonbuild.com}.

\bibitem{Ninja}
E.~Martin, \emph{{Ninja}},
  \href{https://ninja-build.org}{https://ninja-build.org}.

\bibitem{Autotools}
{GNU Project}, \emph{{Autotools}},
  \href{https://www.gnu.org/software/automake/manual/html\_node/Autotools-Introduction.html}{https://www.gnu.org/software/automake/manual/html\_node/Autotools-Introduction.html}.

\bibitem{GiNaC}
C.~Bauer, A.~Frink, R.~B. Kreckel et~al., \emph{{GiNaC is not a CAS}},
  \href{https://www.ginac.de}{https://www.ginac.de}.

\bibitem{Bauer:2000cp}
C.~Bauer, A.~Frink and R.~Kreckel, \emph{{Introduction to the GiNaC Framework
  for Symbolic Computation within the C++ Programming Language}},
  \href{https://doi.org/10.1006/jsco.2001.0494}{\emph{J. Symb. Comput.}
  {\bfseries 33} (2002) 1} [\href{https://arxiv.org/abs/cs/0004015}{{\ttfamily
  cs/0004015}}].

\bibitem{Vollinga:2005pk}
J.~Vollinga, \emph{{GiNaC: Symbolic computation with C++}},
  \href{https://doi.org/10.1016/j.nima.2005.11.155}{\emph{Nucl. Instrum. Meth.}
  {\bfseries A559} (2006) 282}
  [\href{https://arxiv.org/abs/hep-ph/0510057}{{\ttfamily hep-ph/0510057}}].

\bibitem{CLN}
B.~Haible and R.~B. Kreckel, \emph{{CLN - Class Library for Numbers}},
  \href{https://www.ginac.de/CLN}{https://www.ginac.de/CLN}.

\bibitem{ZLIB}
J.-l. Gailly and M.~Adler, \emph{{zlib}},
  \href{https://zlib.net}{https://zlib.net}.

\bibitem{GMP}
{GNU Project}, \emph{{GNU Multiple Precision Arithmetic Library}},
  \href{https://gmplib.org}{https://gmplib.org}.

\bibitem{MPFR}
{GNU Project}, \emph{{GNU Multiple Precision Floating-Point Reliable Library}},
   \href{https://www.mpfr.org}{https://www.mpfr.org}.

\bibitem{YAMLCPP}
J.~Beder et~al., \emph{{yaml-cpp}},
  \href{https://github.com/jbeder/yaml-cpp}{https://github.com/jbeder/yaml-cpp}.

\bibitem{flint}
W.~Hart et~al., \emph{{FLINT: Fast Library for Number Theory}},
  \href{http://www.flintlib.org}{http://www.flintlib.org}.

\bibitem{openmpi}
E.~Gabriel et~al., \emph{{Open Source High Performance Computing}},
  \href{https://www.open-mpi.org}{https://www.open-mpi.org}.

\bibitem{Gabriel:2004}
E.~Gabriel et~al., \emph{{Open MPI: Goals, Concept, and Design of a Next
  Generation MPI Implementation}},
  \href{https://doi.org/10.1007/978-3-540-30218-6_19}{\emph{Adv. Parallel
  Virtual Machine Message Passing Interface. EuroPVM/MPI} {\bfseries 2004}
  (2004) 97}.

\bibitem{mpich}
W.~Gropp et~al., \emph{{MPICH}},
  \href{https://www.mpich.org}{https://www.mpich.org}.

\bibitem{Ellis:2016jkw}
J.~P. Ellis, \emph{{Ti\textit{k}z-Feynman: Feynman diagrams with
  Ti\textit{k}z}},
  \href{https://doi.org/10.1016/j.cpc.2016.08.019}{\emph{Comput. Phys. Commun.}
  {\bfseries 210} (2017) 103}
  [\href{https://arxiv.org/abs/1601.05437}{{\ttfamily 1601.05437}}].

\end{thebibliography}\endgroup

\end{document}